\documentclass[trackchanges,twocolumn]{aastex701}
\usepackage{amsmath}
\usepackage{multirow,booktabs,amssymb,xcolor,siunitx,makecell,rotating,xspace}

\graphicspath{{./}}
\usepackage{cancel}
\usepackage[normalem]{ulem}

\defcitealias{Zuluaga2015}{ZKSA2015}
\defcitealias{Kipping2014}{K2014}
\defcitealias{Seager2003}{SMO2003}

\newcommand{\PhotoRing}{\texttt{PhotoRing}}
\newcommand{\photofit}{PhotoFIT\xspace}

\newcommand{\fullparameters}{p, f_i, f_e, i_R, \theta_R, \tau}
\newcommand{\parameters}{f_e, i_R, \theta_R, p, \tau}
\newcommand{\fullnuisance}{P, \rho_{\star,\mathrm{true}}, b}
\newcommand{\nuisance}{\rho_{\star,\mathrm{true}}, b}
\newcommand{\fullkdeobservables}{\rho_{\star,\mathrm{obs}}, \delta, T_{14}, T_{23}}
\newcommand{\observables}{\rho_{\star,\mathrm{obs}}, \delta, T_{14}, T_{23}, b_{\mathrm{obs}}, (a/R_{\star})_{\rm obs}}
\newcommand{\predictables}{\rho_{\star,\mathrm{obs}}, \delta^\mathrm{geo}, T_{14}^\mathrm{geo}, T_{23}^\mathrm{geo}, b_{\mathrm{obs}}}

\newcommand{\exoplanet}[2]{#1#2}

\definecolor{jamm}{rgb}{0.018, 0.388, 0.878}

\definecolor{numpaque}{rgb}{0.018, 0.388, 0.378} %

\definecolor{attention}{rgb}{1,0,0} %
\makeatletter
\UL@protected\def\PRsout{\leavevmode\bgroup\ULdepth=-.55ex\ULset}
\makeatother
\definecolor{cutcolor}{rgb}{0.45,0.45,0.45}
\DeclareRobustCommand{\cut}[1]{\textcolor{cutcolor}{\PRsout{#1}}}
\renewcommand{\cut}[1]{}

\begin{document}
\title{Probing Exoplanetary Rings with Asterodensity Profiling: A PhotoRing Analysis of Kepler-51}

\author[0000-0002-6140-3116]{Jorge I. Zuluaga}
\affiliation{SEAP/FACom, Instituto de F\'isica - FCEN, Universidad de Antioquia, Calle 70 No. 52-21, Medell\'in, Colombia.}
\email{jorge.zuluaga@udea.edu.co}  

\author[0009-0000-5697-3416]{Sebasti\'an Numpaque}
\affiliation{SEAP/FACom, Instituto de F\'isica - FCEN, Universidad de Antioquia, Calle 70 No. 52-21, Medell\'in, Colombia.}
\email{david.rodriguez1@udea.edu.co} 

\author[0000-0002-4365-7366]{David Kipping}
\affiliation{Columbia University, 550 W 120th Street, New York NY 10027}
\email{dkipping@astro.columbia.edu} 

\author[0000-0003-0353-9741]{Jaime A. Alvarado-Montes} \altaffiliation{Macquarie University Research Fellow (MQRF)} 
\affiliation{Australian Astronomical Optics, Macquarie University, Balaclava Road, Sydney, NSW 2109, Australia.}
\affiliation{Astrophysics and Space Technologies Research Centre, Macquarie University, Balaclava Road, Sydney, NSW 2109, Australia.}
\email{jaime.alvaradomontes@mq.edu.au}

\begin{abstract}
The Kepler-51 system hosts transiting super-puff planets whose extremely low bulk densities challenge structural models, possibly because unresolved rings inflate their inferred radii. We test this hypothesis by combining asterodensity profiling with TTV-aware transit fits of \textit{Kepler} long-cadence photometry. For Kepler-51b we measure a density ratio $\Psi\equiv\rho_{\star,\mathrm{obs}}/\rho_{\star,\mathrm{true}} = 1.18\pm0.06$, a $2.9\sigma$ offset from $\Psi=1$; for Kepler-51d we find $\Psi=1.04\pm0.08$ ($0.6\sigma$), consistent with no anomaly. Unmodelled eccentricity is disfavoured for planet~b by TTV-derived dynamical constraints ($p=0.007$) but remains viable for planet~d ($p=0.33$). The offset of Kepler-51b is also compatible with the \texttt{PhotoRing} (PR) effect of unmodeled rings. Using a geometric ringed-transit forward model and Bayesian retrieval, we identify ring configurations that reproduce the transit observables and the measured density ratios. The preferred solutions are consistent with compact, moderately inclined rings that raise the true bulk densities by factors of 2--5. Conversely, the \emph{absence} of a large PR offset excludes extended, optically thick, high-obliquity rings at $99\%$ credibility for~b and $70\%$ for~d from archival photometry alone. These objects may thus be ``puff ringed planets'' although recent JWST observations of Kepler-51d show no direct signature of an opaque ring. While the required impact parameters and stellar densities show tension with independent constraints---attributable to Kepler's 30-minute cadence---this study presents the first complete application of the PR effect, offered as a methodological demonstration and exclusion experiment rather than as evidence that rings are present. All code is publicly available to support applications to other systems.
\end{abstract}


\section{Introduction} 

Several findings support the hypothesis that planetary rings are not exclusive to our Solar System but a natural outcome of possibly all planetary systems. In the Solar System, all four giant planets possess ring systems \citep{Charnoz2009} linked to tidal disruption of satellites, collisions among moonlets, or residual debris from planetary formation. More recently, the confirmation of rings around small bodies such as Chariklo \citep{Braga-Ribas2014}, Quaoar \citep{Morgado2023}, Chiron \citep{Ortiz2023}, and the dwarf planet Haumea \citep{Ortiz2017} suggests that ring formation can occur under a wide variety of physical conditions, raising more questions about how planetary rings are born and how they evolve. Still, exoplanetary rings (`exorings' for short) remain undiscovered among the thousands of confirmed extrasolar planets. Their detection represents a unique opportunity to explore and provide new constraints on formation and evolution of planetary systems. 

Early theoretical investigations demonstrated that Saturn-like ring systems could produce detectable signatures ($\leq 100$ ppm) in transit photometry \citep{Barnes2004}, especially during ingress and egress. However, although systematic searches for these signals in {\it Kepler} and {\it TESS} data have been conducted over the years \citep{Heising2015,Aizawa2018, Umetani2025RingsTESS}, no unambiguous detections of rings around transiting exoplanets have yet been confirmed \citep{Brown2001, Santos2015, Osborn2017, Ballesteros2018, LecavelierdesEtangs2017, Akinsanmi2020}. In practice, an obscuring structure around 1SWASP J140747b is presently considered the only confirmed exoring case \citep{Mamajek2012, Kenworthy2015}. However, given how young the system is, these ``rings'' are most likely composed of dust and gas and probably represent the remains of a circumplanetary disc, rather than debris from previous tidal interactions or moon disruptions. 

Exorings could also be detected via modern high-contrast imaging techniques \citep{Follette2023} or reflected-light photometry \citep{Arnold2004,Sucerquia2020,Zuluaga2022} and polarimetry \citep{Veenstra2025}. However, because these techniques are still limited, most studies have concentrated on identifying their distinctive imprints in stellar light curves, which change according to their orientation, radial extent, and optical depth (see eg. \citealt{Ohta2009,Zuluaga2015}).

The most salient effect of an exoring on a transit light-curve arises from the fact that the occulting area of a ringless planet can be significantly smaller than that of a ringed planet. If rings are not included in the transit model of the latter, the planetary radius will be overestimated (see e.g., \citealt{Zuluaga2015}). This would result in the bulk density being underestimated, which in turn could cause the planet to be incorrectly identified as an unusually low-density, “puffed” planet \citep{bakos2007hat}. In extreme cases, exorings could explain the anomalous densities of the so-called ``super-puffs'' or super puff planets  ($\rho<0.1$ g/cm$^3$, see eg. \citealt{Lee2016}), bodies with radii comparable to those of gas giants but only a few times the mass of Earth.

The planetary system Kepler-51 contains some of the most striking examples of so‑called super-puff planets. This compact, multi-planet system hosts at least three transiting planets—\exoplanet{Kepler-51}{b}, \exoplanet{Kepler-51}{c}, and \exoplanet{Kepler-51}{d}—with sizes between those of Neptune and Saturn, whose masses have been inferred from transit-timing variation (TTV) analyses (\citealt{Masuda2014}). With estimated masses of 5–6 $M_\oplus$, these planets are formally categorized as super-Earths. Their radii, however, are on the order of 6–10 $R_\oplus$, comparable to gas giants, which results in extremely low mean densities of $\rho < 0.06$ g cm$^{-3}$. As a consequence, they are currently the least dense—and thus the most extremely inflated—planets known. More recently, a fourth planet, Kepler-51 e, has been suggested \citep{Masuda2024}. Its presence is inferred indirectly, and its characteristics are consequently less well determined than those of the other planets in the system.

Several physical processes have been suggested to explain the extremely low densities and inflated radii of the Kepler-51 planets, including extended H-He envelopes, atmospheric escape and the possibility that they are still at a relatively early evolutionary stage (see, e.g., \citealt{Libby-Roberts2020}; \citealt{Wang2019}). Recent JWST observations of Kepler-51d, however, show, on the one hand, no significant photometric signature indicative of an equatorial bulge or a large opaque ring system \citep{Lammers2024}, and, on the other hand, a sloped transmission spectrum \citep{Libby-Roberts2025} that strongly supports the presence of a high-altitude haze layer, rendering the existence of a massive, opaque exoring highly unlikely. While an optically thin, very porous ring around Kepler-51d cannot be entirely excluded, prior theoretical work has already shown that standard rocky rings struggle to account for the extreme sizes of the Kepler-51 planets without assuming exceptionally porous material \citep{piro2020exploring}. In turn, the true nature of Kepler-51b remains unclear without further follow-up observations. Owing to this ongoing debate and their peculiar physical characteristics, the Kepler-51 planets serve as an excellent testing ground for the development and implementation of novel methods.

In \citet{Zuluaga2015} (hereafter \citetalias{Zuluaga2015}), the so‑called \PhotoRing\ (PR) effect was introduced, a technique for identifying planetary rings even when transit light curves lack prominent, well‑resolved morphological signatures. PR is part of the broader class of asterodensity profiling (AP) techniques (\citealt{Kipping2014}, hereafter \citetalias{Kipping2014}). In AP, the mean stellar density of the host star inferred from transit observables (the “observed density”)—namely the transit duration, depth, impact parameter, and orbital period— is compared with the bulk stellar density obtained from independent stellar characterization methods (the “true density”), such as asteroseismology or stellar isochrone-fitting. When the transit signal is influenced by unmodeled phenomena, such as stellar blending, orbital eccentricity, or the presence of planetary rings, the observed density will differ from the true density.

Despite the potential of the PR effect and the community’s positive reception of the concept, the method has not yet been applied to real planetary systems (see eg. \citealt{heising2015search, piro2020exploring, millholland2025exploring, umetani2025search}). With over a decade of accumulated observations, Kepler-51 now emerges as an optimal target to test the method and help unravel the baffling properties of its planets. To this end, we employ here PR to investigate whether the anomalous characteristics of the Kepler-51 planets can at least be partially explained due to the presence of planetary rings. We infer a stellar density from the Kepler transit light curves—among other transit observables—for each planet and compare it with an independent, isochrone-based estimate of the true stellar density \citep{Masuda2024}, searching for discrepancies indicative of the PR effect. We then applied the geometric forward model for transiting ringed planets of \citetalias{Zuluaga2015} to predict the effective transit observables—namely, the transit depth $\delta$ and the durations $T_{14}$ and $T_{23}$—as well as the corresponding $\rho_{\star,\mathrm{obs}}$ for a specified ring configuration. 

This paper is organized as follows. Section~\ref{sec:data_analysis} presents the observational data analysis, including our light-curve fit of the Kepler-51 planets. Section~\ref{sec:anomalous_stellar_densities} compares the resulting transit-inferred stellar densities with independent isochrone-based estimates, and Section~\ref{sec:photoeccentric_effect} assesses whether unmodelled orbital eccentricity can account for the measured offsets. Section~\ref{sec:geom_model} introduces the geometric forward model used to predict effective transit observables for ringed planets. Section~\ref{sec:asterodensity_profiling} then sets out the asterodensity profiling framework and the \PhotoRing\ effect. Section~\ref{sec:bayes_inference} details the Bayesian inference methodology, including the likelihood construction and the treatment of nuisance parameters. Section~\ref{sec:ring_results} reports the results of our retrievals, detailing the inferred ring geometries and evaluating the model through posterior predictive checks. Section~\ref{sec:discussion} discusses the physical implications of our findings and presents our concluding remarks.

\section{Observational Data Analysis}
\label{sec:data_analysis}

The observational and data basis of this study is grounded in the posterior distributions of the asterodensity profiling observables. ($\delta, \rho_{\star,\mathrm{obs}},  b_{\rm obs}$) for \exoplanet{Kepler-51}{b} and \exoplanet{Kepler-51}{d}, the innermost and outermost planets of Kepler-51, obtained from a dedicated light-curve fit of the available \textit{Kepler} photometry (see \autoref{sec:data_analysis_ttv}). These posteriors are not derived from a direct ring-transit model, they rather represent the inferences made under the standard assumption of a spherical, ringless planet, thus encoding the effective transit observables that must result from any ringed-planet configuration.

In this section, we describe the origin of these posteriors and how they are incorporated into our ring-retrieval framework. Although more recent observations of the system have been obtained with JWST, HST, and TESS facilities \citep{Masuda2024, Lammers2024, Libby-Roberts2025}, the present analysis exclusively relies on constraints derived from the original \textit{Kepler} photometry, without including the aforementioned new datasets. This restriction is driven by two primary factors: first, to assess how effectively our analysis pipeline operates when only a limited amount of photometric data is available; and second, because our team did not have access to the full data set for most of the duration of this study. Although we could, in principle, wait to obtain all existing photometric measurements of the system, we will demonstrate below that the main limitation on the use of AP effects is the precise determination of $\rho_{\star,true}$, which has not been significantly improved by the more recent observations of the system. Nevertheless, as future measurements yield more accurate estimates of this parameter, the methods and tools presented here will prove highly valuable for addressing the puzzle at hand, particularly that posed by planet \exoplanet{Kepler-51}{b} (see \autoref{sec:photoeccentric_effect}).

\subsection{Transit and \photofit Modeling}
\label{sec:data_analysis_ttv}

For a robust inference of the transit parameters in the presence of TTVs, and without resorting to a full dynamical analysis, we employed a so-called {\em photometric fit with independent transit times} (hereafter \photofit). Enforcing a strict linear ephemeris on systems with significant TTVs can artificially smear the transit shape. In our \photofit approach, the central transit time ($T_c$) for each epoch is treated as an independent free parameter during the light curve fitting process. This decouples the transit's photometric shape from the underlying orbital dynamics, allowing us to accurately measure the geometric transit parameters while also extracting the individual transit epochs.

For \exoplanet{Kepler-51}{b} and \exoplanet{Kepler-51}{d}, the asterodensity profiling posterior distributions were sampled using the nested sampling algorithm \texttt{MultiNest} \citep{Feroz2009, Feroz2019} applied to the \textit{Kepler} photometric time series\footnote{We exclude the recently discovered fourth planet, Kepler-51 e, and \exoplanet{Kepler-51}{c} due to its grazing transit geometry \citep{Masuda2024}}. The analysis fits each transit simultaneously with our \photofit model, extracting individual mid-transit times for each observed epoch while allowing to estimate the AP parameters from the full ensemble of transits.

For \exoplanet{Kepler-51}{d}, the small number of observed transits and the corresponding parameter space allowed a single nested-sampling run, yielding one posterior distribution per parameter. For \exoplanet{Kepler-51}{b}, however, the large number of observed epochs translates into a high-dimensional parameter space---since each epoch contributes an individual mid-transit time as a free parameter---making a single nested-sampling run computationally intractable. To handle this high dimensionality, the full photometric time series was divided into three independent temporal segments, each covering a distinct subset of observed transits. A separate run was performed on each segment, producing its corresponding posterior distributions for AP parameters. We then  combine the resulting distributions into a single posterior by running a short Markov Chain Monte Carlo sampler that simultaneously draws from the three posterior distributions, treating them as statistically independent constraints on the transit parameters. The resulting chain defines the final posterior of \exoplanet{Kepler-51}{b} used throughout this work.

\subsection{Derived Observables}
\label{sec:derived_observables}

For each planet, the main outputs of the \photofit are joint posterior samples of the asterodensity profiling (AP, see \autoref{sec:asterodensity_profiling}) parameters $(\rho_{\star,\mathrm{obs}}, b_{\mathrm{obs}}, \delta)$, along with the orbital period $P$ and the limb-darkening coefficients. For clarity, \autoref{tab:symbols} compiles the definitions of all quantities employed in this work. The meaning of each quantity and its function in the analysis are explained in the relevant sections (e.g., the ringed-planet parameters of the forward model are introduced in \autoref{sec:geom_model}, the properties obtained from asterodensity profiling are specified in \autoref{sec:photoeccentric_effect}, and so on). 

\begin{figure*}[t]
\centering
\includegraphics[width=\textwidth]{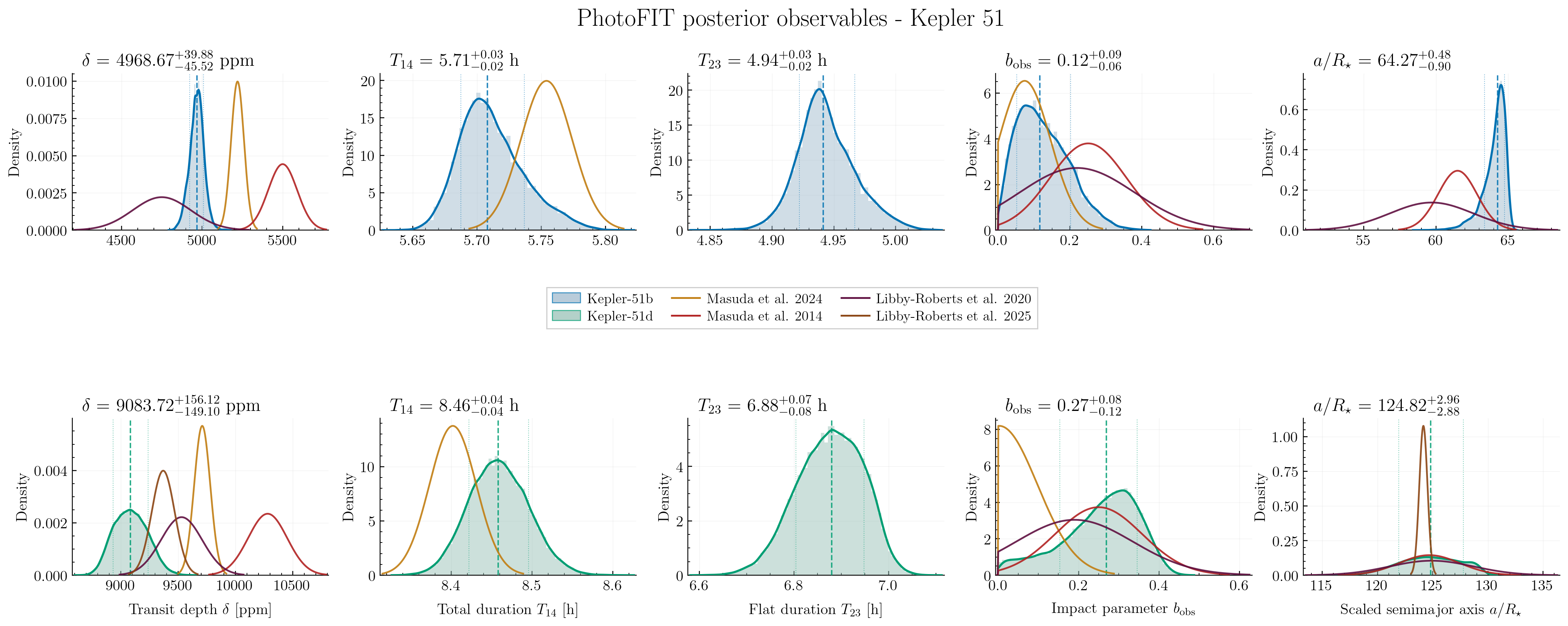}
\caption{Marginal posterior distributions of the observables for \exoplanet{Kepler-51}{b} (top row) and \exoplanet{Kepler-51}{d} (bottom row). Those obtained from the \photofit carried out in this work are shown as filled distributions. Vertical dashed lines and the accompanying labels indicate the median and the 68\% credible interval for each marginal posterior. For comparison, literature estimates are depicted as normal distributions (unfilled curves) constructed from the published point values and their reported uncertainties; in the case of \citet{Masuda2024}, a truncated normal distribution is used for the impact parameter $b$ to respect its physical limits. Only those observables provided by each reference are plotted (for example, none of the literature sources report values for the flat-bottom duration $T_{23}$).}
\label{fig:observables_full}
\end{figure*}

Assuming a ringless planet, these parameters capture all of the transit-geometry information encoded in the \textit{Kepler} light curves. The secondary transit quantities $(a/R_\star)_{\mathrm{obs}}, T_{14}$, and $T_{23}$, which are needed for our AP analysis, are then deterministically obtained by inverting the relations given in \citetalias{Kipping2014}. Their posterior distributions are derived directly from the same posterior samples, ensuring that the full covariance structure inferred from the photometry is preserved. In \autoref{fig:observables_full}, we show the marginal posterior distributions for the full set of observables $(\observables)$ for both \exoplanet{Kepler-51}{b} and \exoplanet{Kepler-51}{d}. Our results (shaded curves) are compared with those of other authors (see next section).

\begin{table}[t]
\centering
\small
\renewcommand{\arraystretch}{1.2}
\caption{Summary of symbols and notation used throughout this work.}
\label{tab:symbols}
\begin{tabular}{ll}
\hline
\hline
Symbol & Description \\
\hline
\multicolumn{2}{l}{\textit{Nuisance parameters of the forward model}} \\
$P$ & Orbital period [hours] \\
$b$ & Transit impact parameter [--] \\
$\rho_{\star,\mathrm{true}}$ & Independent stellar density [g\,cm$^{-3}$] \\
\hline
\multicolumn{2}{l}{\textit{Ringed planet parameters of the forward model}} \\
$p$ & $R_p/R_\star$, scaled true planetary radius [--] \\
$f_i, f_e$ & Inner and outer ring radius [$p$] \\
$i_R, \theta_R$ & Projected ring inclination and tilt [deg] \\
$\alpha$ & $\exp(-\tau)$, Ring normal attenuation [--] \\
$\tau$ & Ring normal opacity [--] \\
\hline
\multicolumn{2}{l}{\textit{Derived parameters of the forward model}} \\
$(a/R_{\star})_\mathrm{geo}$ & Geometric scaled semi-major axis[--] \\
$i_\mathrm{orb}$ & Orbital inclination [deg] \\
$\delta^\mathrm{geo}$ & Geometric transit depth [ppm] \\
$T_{14}^\mathrm{geo}$ & Geometric total transit duration [hours] \\
$T_{23}^\mathrm{geo}$ & Geometric full transit duration [hours] \\
\hline
\multicolumn{2}{l}{\textit{Transit (photometric) observables}} \\
$\delta$ & Transit depth [ppm] \\
$T_{14}$ & Total transit duration [hours] \\
$T_{23}$ & Full-transit duration [hours] \\
$p_\mathrm{obs}$ & Ringless-model scaled planetary radius [--] \\
\hline
\multicolumn{2}{l}{\it Asterodensity profiling derived properties} \\
$(a/R_{\star})_\mathrm{obs}$ & Scaled semi-major axis [--] \\
$\rho_{\star,\mathrm{obs}}$ & Transit-inferred stellar density [g\,cm$^{-3}$] \\
$\Psi$ & $\rho_{\star,\mathrm{obs}}/\rho_{\star,\mathrm{true}}$, density anomaly [--] \\
PR & $10\,\log_{10}\Psi$, logarithmic PR scale [--] \\
$\Theta_{ij}$ & $(\Psi_i/\Psi_j)^{2/3}$, Ratio of density anomalies [--]\\
&  (multiplanet asterodensity profiling, MAP) \\
\multicolumn{2}{l}{\textit{Bayesian inference}} \\
\hline
$\boldsymbol{\theta}$ & Ringed planet parameter vector  \\
$\boldsymbol{\eta}$ & Nuisance parameters  \\
$\mathbf{y}$ & Observable vector  \\
$\mathcal{L}$ & Likelihood function  \\
$\hat{p}(\mathbf{y})$ & KDE estimate of \photofit posterior  \\
\hline
\hline
\end{tabular}
\end{table}

For a specified set of ring parameters $(p, f_i, f_e, i_R, \theta_R, \alpha)$ and suitable choices of orbital and stellar properties $(P, b, \rho_\mathrm{\star,true})$, the geometrical forward model predicts the corresponding observable vector. The Bayesian inference on the ring parameters is subsequently carried out by confronting this model prediction with the distribution derived from the light curves. This procedure represents the foundation of the PR analysis.

\subsection{Literature Comparison}
\label{sec:data_analysis_comparison}

The Kepler-51 system has been characterized by several independent photometric analyses spanning a decade: the confirmed three-planet architecture by \citet{Masuda2014}, the atmospheric characterization of \citet{Libby-Roberts2020} and \citet{Libby-Roberts2025}, and the refined dynamically re-analysis and discovery of a fourth non-transiting planet, \exoplanet{Kepler-51}{e} of \citet{Masuda2024}. Since our goal is to obtain posterior distributions of the photometric transit observables that enter the asterodensity profiling and ring inference described later in this work, the comparison between our \photofit-derived posteriors and the literature values therefore serves two purposes: (i) to validate that our analysis recovers transit shapes consistent with independent measurements, and (ii) to identify and explain genuine discrepancies that may arise from differences in data selection, dynamical assumptions, or statistical frameworks. In \autoref{tab:observables_full}, we summarize our \photofit transit observables along with the corresponding literature values.

A visual comparison of the values reported in the literature with those derived in this work is presented in \autoref{fig:observables_full}. Because the posterior distributions of the observables reported by other authors are not available, we assume in all cases that they follow Gaussian distributions (unfilled curves). While our own posteriors demonstrate that observable distributions can substantially deviate from normality, this assumption is made solely for comparative purposes and our inference pipeline does not use the exact shapes of the literature distributions.

\begin{table*}[t]
\footnotesize
\centering
\renewcommand{\arraystretch}{1.2}
\setlength{\tabcolsep}{15pt} 
\caption{Transit observables for \exoplanet{Kepler-51}{b} and \exoplanet{Kepler-51}{d} from this work (\photofit posterior) and literature values.}
\label{tab:observables_full}
\begin{tabular}{lccl}
\hline\hline
Observable & \exoplanet{Kepler-51}{b} & \exoplanet{Kepler-51}{d} & Source \\
\hline
\multirow{5}{*}{$p$ [$R_{\star}$ ($R_{\oplus}$)]} & $0.07055_{-0.00030}^{+0.00026}\,(6.687_{-0.028}^{+0.024})$ & $0.09531_{-0.00078}^{+0.00082}\,(9.034_{-0.074}^{+0.078})$ & This work (\photofit) \\
 & $0.07225 \pm 0.0003 \,(6.83 \pm 0.13)$ & $0.09857 \pm 0.00037\,(9.32 \pm 0.18)$ & \citet{Masuda2024} \\
 & $0.07414_{-0.00061}^{0.00059}\,  (7.1\pm 0.3)$ & $0.10141_{-0.00085}^{0.00084} \, (9.7 \pm 0.5)$ & \citet{Masuda2014} \\
 & $0.0689\pm0.0013\,(6.89\pm0.14)$ & $0.0976\pm0.0009\,(9.46\pm0.16)$ & \citet{Libby-Roberts2020} \\
 & -- & $0.09682 \pm 0.00051\, (9.19 \pm 0.05)$ & \citet{Libby-Roberts2025} \\
\hline
\multirow{5}{*}{$\delta$ [ppm]} & $4977_{-42.1}^{+36.1}$ & $9084_{-149}^{+156}$ & This work (\photofit) \\
 & $5220 \pm 40$ & $9710 \pm 70$ & \citet{Masuda2024} \\
 & $5500 \pm 90$ & $10280 \pm 170$ & \citet{Masuda2014} \\
 & $4750 \pm 180$ & $9530 \pm 180$ & \citet{Libby-Roberts2020} \\
 & -- & $9370 \pm 100$ & \citet{Libby-Roberts2025} \\
\hline
\multirow{2}{*}{$T_{14}$ [h]} & $5.708_{-0.0176}^{+0.0238}$ & $8.458_{-0.036}^{+0.0377}$ & This work (\photofit) \\
 & $5.754 \pm 0.02$ & $8.402 \pm 0.029$ & \citet{Masuda2024} \\
\hline
\multirow{1}{*}{$T_{23}$ [h]} & $4.939_{-0.019}^{+0.0243}$ & $6.88_{-0.0752}^{+0.0686}$ & This work (\photofit) \\
\hline
\multirow{4}{*}{$b_{\rm obs}$} & $0.1443_{-0.0579}^{+0.0551}$ & $0.2696_{-0.116}^{+0.0757}$ & This work (\photofit) \\
 & $0.074 \pm 0.072$ & $0.003 \pm 0.095$ & \citet{Masuda2024} \\
 & $0.251 \pm 0.106$ & $0.25 \pm 0.108$ & \citet{Masuda2014} \\
 & $0.22 \pm 0.16$ & $0.19 \pm 0.145$ & \citet{Libby-Roberts2020} \\
\hline
\multirow{4}{*}{$a/R_{\star}$} & $64.08_{-0.647}^{+0.46}$ & $124.8_{-2.88}^{+2.96}$ & This work (\photofit) \\
 & $61.5 \pm 1.35$ & $124.7 \pm 2.75$ & \citet{Masuda2024} \\
 & $59.7 \pm 2.9$ & $124.9 \pm  3.8$ & \citet{Libby-Roberts2020} \\
 & -- & $124.2 \pm 0.37$ & \citet{Libby-Roberts2025} \\
\hline
\end{tabular}
\end{table*}

The transit depth $\delta$ is the most straightforward quantity to determine from the data. \autoref{tab:observables_full} and \autoref{fig:observables_full} reveal a pronounced scatter among published values for both planets that is larger than the quoted uncertainties of the individual measurements. The most natural explanation for this spread is \emph{unocculted stellar activity}, rather than any systematic issue in the light-curve fitting. This phenomenon was explicitly detected in the follow-up observations of \exoplanet{Kepler-51}{d} using NIRspec, as reported by \citet{Lammers2024}. Kepler-51 is a young ($\sim$500\,Myr), magnetically active star whose evolving starspot distribution alters the apparent out-of-transit flux level, artificially increasing the measured transit depth if the spot contribution is not explicitly modeled. Since the spot covering fraction changes over the \textit{Kepler} observing window, and our analysis does not apply a dedicated correction for stellar spots, the derived depths should be interpreted as an average over the star’s activity state across the different epochs. This explains their agreement with the full span of literature values and supports the conclusion that the observed dispersion is predominantly caused by time-varying stellar activity, rather than by differences in analysis methodology.

On the other hand, the impact parameter $b_{\mathrm{obs}}$ is among the least constrained transit observables reported in the literature, with published estimates spanning a wide range for both planets. This behavior arises primarily from the limited temporal resolution of the \textit{Kepler} long-cadence data ($\sim$30 min), which is comparable to the ingress/egress durations of the planets. Consequently, this averages and smooths the slopes of those regions that contain most of the information used to constrain the transit chord geometry and, therefore, the impact parameter. This loss of geometric information makes $b_{\rm obs}$ highly degenerate, which can only be effectively broken with higher-cadence observations---as in the JWST analysis of \citet{Masuda2024} that resolves a nearly central value for \exoplanet{Kepler-51}{d} ($b \sim 0.003$). However, they also show that even with high-quality data, the solution for $b_{\rm obs}$ of \exoplanet{Kepler-51}{b} remains bimodal, thus implying the existence of two distinct transit geometries that fit the observations equally well. Our posterior for $b_{\rm obs}$, obtained from a joint fit of all available epochs, is consistent with both the low-$b$ and high-$b$ solutions reported and shows no significant tension with individual measurements. This reflects the degeneracy imposed by the temporal resolution of the available photometry instead of a physical disagreement.

The total transit duration $T_{14}$ and the flat-bottom duration $T_{23}$ are not frequently tabulated in the literature, because most works describe the transit shape directly using $(a/R_\star)_{\mathrm{obs}}$ and $b_{\mathrm{obs}}$ rather than the contact-time durations themselves. In the one case where $T_{14}$ has been published by \citet{Masuda2024}, our posterior values for both planets are consistent, differing by at most $\lesssim4$ minutes, which lies comfortably within the quoted uncertainties. In contrast, the scaled semi-major axis $(a/R_\star)_{\mathrm{obs}}$ is the most consistently determined observable across all datasets and analyses: our measurements for both planets agree with those of \citet{Masuda2024}, \citet{Libby-Roberts2020}, and the independent JWST-based determination of \citet{Libby-Roberts2025} to within $1\sigma$. This agreement shows that the orbital scale of the system is effectively unaffected by the cadence, wavelength coverage, or modeling assumptions described above, and that $(a/R_\star)_{\mathrm{obs}}$ is set almost entirely by the precisely measured orbital period and total transit duration.

\section{Transit-inferred versus Independent Stellar Densities}
\label{sec:anomalous_stellar_densities}

A key consequence of comparing transit-inferred and independently determined stellar densities \citepalias{Kipping2014} is that, within the standard transit model—assuming unblended, spot-free light curves and spherical planets on circular orbits—all transiting planets in the same system must yield an identical value of $\rho_{\star,\mathrm{obs}}$. This conclusion follows directly from Kepler's third law: the inferred stellar density is computed from the measured orbital scale $(a/R_\star)_{\mathrm{obs}}$ that only depends on the transit light-curve morphology and the orbital period (see \autoref{sec:asterodensity_profiling} for more details). Consequently, if $\rho_{\star,\mathrm{obs}}$ differs among multiple planets that orbit the same star, this indicates that the standard transit model is failing to capture some aspect of the actual transit geometry for at least one of the planets, or that the light curves are being systematically modified by a physical effect not accounted for in the model. 

\begin{figure*}[t]
\centering
\includegraphics[width=\textwidth]{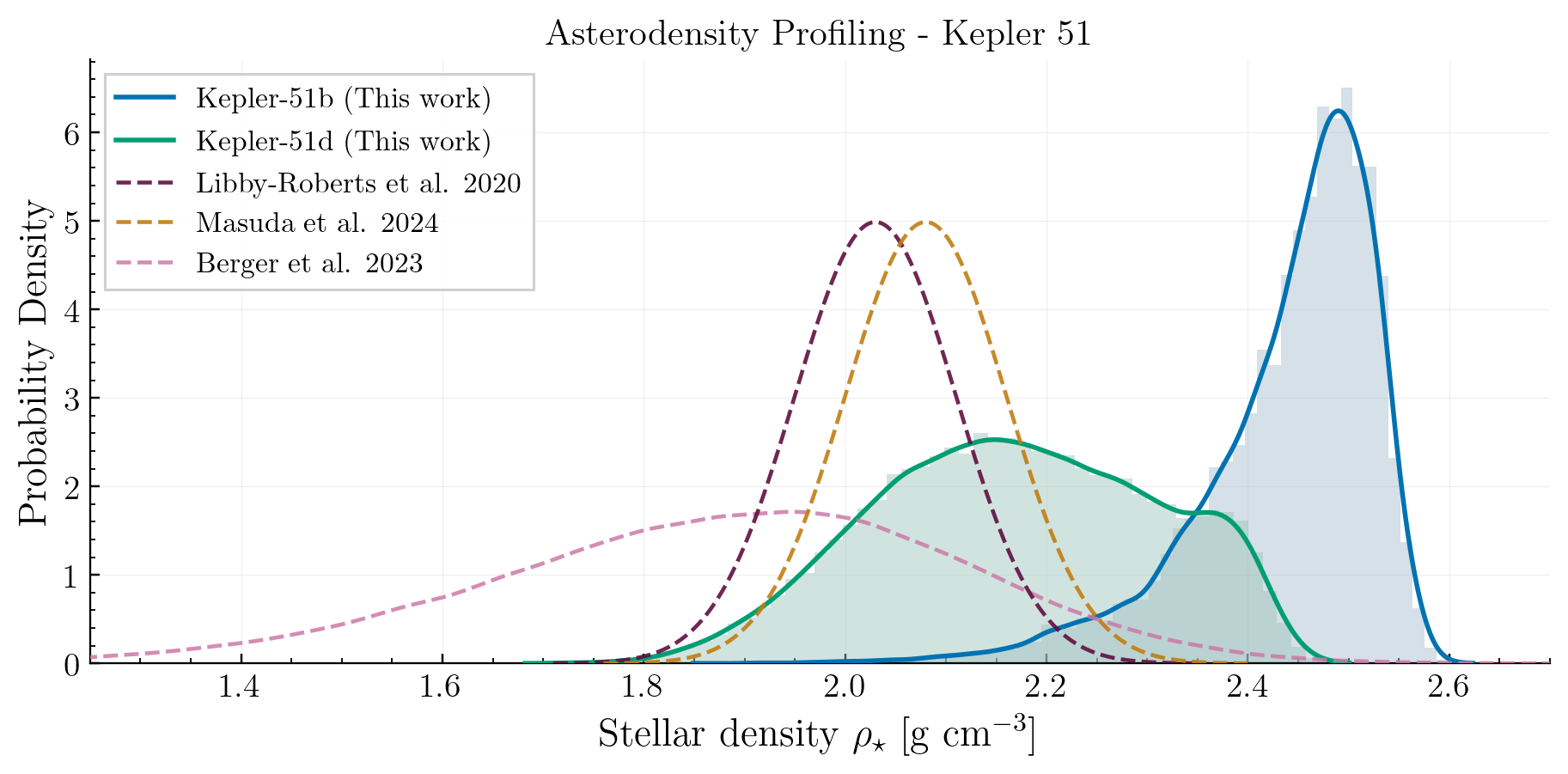}
\caption{Marginal posterior distributions of the transit-derived stellar density, $\rho_{\star,\mathrm{obs}}$, for \exoplanet{Kepler-51}{b} (blue) and d (orange), obtained from the \photofit presented here (filled histograms and continuous lines). The discontinuous curves correspond to the value of the true stellar density $\rho_{\star,\mathrm{true}}$ ad obtained by independent works using isochrone-based posterior samples \citep{Libby-Roberts2020, Berger2023, Masuda2024}.} 
\label{fig:rhoobs_comparison}
\end{figure*}

\autoref{fig:rhoobs_comparison} presents the marginal posterior distributions of $\rho_{\star,\mathrm{obs}}$ derived from our photometric analysis for both planets (filled histograms and continuous lines), together with the corresponding distribution of $\rho_{\star,\mathrm{true}}$ from the independent isochrone-based estimates of \citet{Berger2023}, \citet{Libby-Roberts2020} and \citet{Masuda2024} (dashed, empty curves). For the latter cases, we represent their posterior distributions as normal distributions, using the reported means and uncertainties. For ease of comparison, the mean values and associated 1-$\sigma$ or 68\% uncertainties from our work and from the literature are compiled in \autoref{tab:rho_star_compendium}. 

\begin{table*}[t]
\centering
\renewcommand{\arraystretch}{1.2}
\caption{Estimates of the stellar density for Kepler-51. The values are grouped according to whether they are obtained from transit observables (including photodynamical modeling) or from independent stellar characterization. The transit-based estimates reported by \citet{Masuda2024} were derived under the assumption of the previously known three-planet configuration, whereas their isochrone-based stellar characterization was carried out using the current four-planet model. For reference, we also list an estimate of the logarithmic PR scale $\mathrm{PR}=10\log_{10}(\rho_{\star}/2\,\mathrm{g\,cm^{-3}})$, computed by adopting a true reference stellar density of $2$ g cm$^{-3}$, which is close to the mean of the independent estimates. The uncertainties on PR were computed via straightforward linear error propagation. The PR values are reported for context only and are not used in the inference analysis. IsoFIT stands for Isochrone fitting.\label{tab:rho_star_compendium}}
\begin{tabular}{llcccl}
\hline\hline
Category & Method & $\rho_{\star}$ [g cm$^{-3}$] & $\Psi$ & PR (ref.) & Source \\
\hline

\multirow{6}{*}{Transit-derived}
& \photofit posterior
& $2.441_{-0.073}^{+0.053}$
& $1.18\pm 0.06$
& $0.87\pm 0.12$
& This work (\exoplanet{Kepler-51}{b})
\\

& \photofit posterior
& $2.171_{-0.147}^{+0.158}$ 
& $1.04\pm 0.08$
& $0.36\pm 0.31$
& This work (\exoplanet{Kepler-51}{d})
\\

& Transit light-curve fit
& $2.16_{-0.13}^{+0.15}$ 
& --
& $0.33\pm 0.28$
& \citet{Masuda2014}
\\

& Transit light-curve fit
& $2.42 \pm 0.06$ 
& --
& $0.83\pm 0.11$
& \citet{Masuda2024}
\\

& N-body Photodynamical fit
& $\sim 2.56_{-0.34686}^{+0.423}$ 
& --
& $1.07\pm 0.65$
& \citet{Masuda2024}
\\

\hline

\multirow{4}{*}{Independent}

& IsoFIT (CKS+Gaia DR2)
& $2.03 \pm 0.08$ 
& --
& --
& \citet{Libby-Roberts2020}
\\

& IsoFIT (MIST+Gaia DR3)
& $2.08 \pm 0.08$ 
& --
& --
& \citet{Masuda2024}
\\

& IsoFIT (Gaia DR3)
& $1.896_{-0.246}^{+0.218}$ 
& --
& --
& \citet{Berger2023}
\\
\hline
\end{tabular}
\end{table*}

To quantify the offset while propagating the uncertainty of both quantities, we compute the density stellar ratio \citep{kipping2012novel},

\begin{equation}
\Psi \equiv \left(\frac{\rho_{\star,\mathrm{obs}}}{\rho_{\star,\mathrm{true}}}\right)
\label{eq:PR_Kipping}
\end{equation}
at the sample level, pairing samples from our \photofit posterior distributions with values taken from a normal distribution defined by $\mu_\rho = 2.08$ and $\sigma_\rho = 0.08$~g\,cm$^{-3}$, matching the isochrone-based (MIST+Gaia DR3) estimate of \citet{Masuda2024}, which was obtained under the currently preferred four-planet configuration of the system.

We obtain $\Psi_b = 1.18 \pm 0.06$ and $\Psi_d = 1.04 \pm 0.08$ (medians and 68\% credible intervals), with 95\% credible intervals of $[1.04, 1.29]$ and $[0.90, 1.20]$. Measured against the no-effect value, $\Psi=1$, \exoplanet{Kepler-51}{b} lies $2.9\sigma$ below the median with $P(\Psi_b \le 1) = 0.8\%$, but only $0.6\sigma$ below it for \exoplanet{Kepler-51}{d}, with $P(\Psi_d \le 1) = 29\%$. With this narrower stellar prior the error budget is now dominated by the photometry rather than by the stellar characterization, most clearly for \exoplanet{Kepler-51}{d}, where $\rho_{\star,\mathrm{obs}}$ carries a $\sim7\%$ uncertainty against $\sim4\%$ on $\rho_{\star,\mathrm{true}}$.

The two planets must therefore be treated independently, and our interpretation of the corresponding values is intentionally conservative. \exoplanet{Kepler-51}{b} exhibits an offset that is unlikely to be explained by random noise alone, but it still does not reach any threshold that would justify a formal detection. In contrast, \exoplanet{Kepler-51}{d} is fully compatible with the absence of any asterodensity profiling signal: its 68\% credible interval, $[0.97, 1.13]$, includes unity. Accordingly, we describe these quantities as density \emph{offsets} rather than as ``anomalies'' in what follows, and we opt to simply report the values and allow the reader to decide whether an effect is present.

Several mechanisms could contribute to any genuine difference between the stellar density inferred from photometry and that obtained from other stellar-astrophysics methods: 

\begin{itemize}
    \item[(i)] An underlying non-circular orbit, commonly known as the photoeccentric effect \citep{Dawson2012,kipping2012novel,Kipping2014}. If a planet transits near periastron, its higher orbital velocity mimics a closer orbit, leading to an artificially inflated inferred stellar density.
    
    \item[(ii)] The omission of an explicit spot correction in the photometric analysis, which affects the measured $(a/R_\star)_{\mathrm{obs}}$ and, in turn, $\rho_{\star,\mathrm{obs}}$. In our study, this effect could partially explain the higher stellar densities derived for both planets, as outlined in \autoref{sec:data_analysis_comparison}. 
    
    \item[(iii)] Unmodeled dynamical interactions among the planets. Kepler-51 exhibits large-amplitude transit timing variations (TTVs) that can distort the effective transit light-curve shape if not properly accounted for. In our analysis, we substantially mitigate this bias by fitting a separate mid-transit time for each individual epoch. More generally, an incomplete characterization of the system’s architecture can also bias stellar-density estimates. This is demonstrated by the three-planet configuration assumed by \citet{Masuda2014}, as well as by both the transit-fitting and photodynamical analyses of \citet{Masuda2024} (\autoref{tab:rho_star_compendium}). That truncated dynamical model incorrectly predicted the JWST-measured mid-transit time of Kepler-51d by \(\sim 120\) minutes. Consequently, stellar-density inferences that rely on this assumption—including those presented here, which likewise neglect perturbations from the fourth planet—may also be affected by systematic biases. 
    
    \item[(iv)] The existence of an extra geometric feature—such as a ring system—that changes the observed transit configuration by enlarging the effective occulting area and altering the transit contact times. 
\end{itemize}

\section{Assessing the Photoeccentric Effect}
\label{sec:photoeccentric_effect}

As explained in point (i) of \autoref{sec:anomalous_stellar_densities}, the positive density offset observed in Kepler-51b and d ($\Psi \equiv \rho_{\star,\mathrm{obs}}/\rho_{\star,\mathrm{true}} > 1$) could in principle be induced by an unmodeled orbital eccentricity if the planets transit near periastron. This is known as the photoeccentric effect \citep{Dawson2012}. To verify whether this mechanism can account for the observed $\rho_{\star,\mathrm{obs}}$, we calculate the minimum eccentricity $e_{\mathrm{min}}$ required by \photofit and compare it against the dynamical constraints imposed by TTVs \citepalias{Kipping2014}. 

For a single planet, the minimum eccentricity capable of producing  $\Psi > 1$ (which occurs when the argument of periastron $\omega = 90^\circ$) is given by:
\begin{equation}
    e_{\mathrm{min}} = \frac{\Psi^{2/3} - 1}{\Psi^{2/3} + 1}.
\end{equation}

\begin{figure}[t]
\centering
\includegraphics[width=\columnwidth]{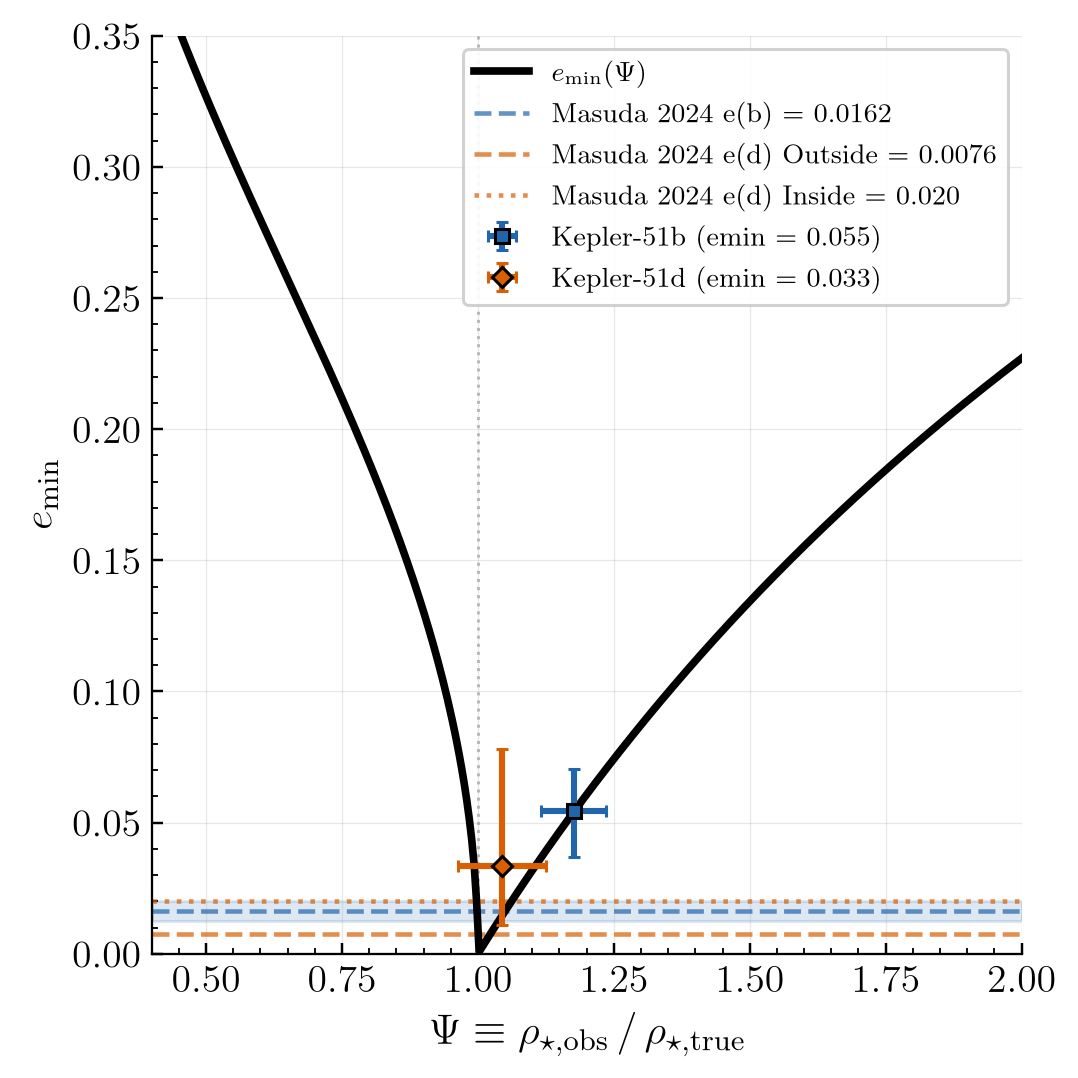}
\caption{The minimum required eccentricity $e_{\mathrm{min}}$ plotted as a function of the density ratio $\Psi = \rho_{\star,\mathrm{obs}}/\rho_{\star,\mathrm{true}}$ (solid black curve). The dashed curves show the actual eccentricity values obtained from the detailed TTV analysis by \citet{Masuda2024}.}
\label{fig:pe_emin_diagnostic}
\end{figure}

By evaluating this expression over our posterior samples of $\Psi$ for each planet, we find $e_{\mathrm{min}} = 0.054_{-0.018}^{+0.016}$ for Kepler-51b $e_{\mathrm{min}} = 0.034_{-0.022}^{+0.044}$ for Kepler-51d. The interval stays relatively wide for \exoplanet{Kepler-51}{d} because $e_{\mathrm{min}}$ depends very sensitively on $\Psi$ when $\Psi$ is near 1, and this planet’s value of the ratio is close to unity. As illustrated in \autoref{fig:pe_emin_diagnostic}, these photometric requirements are larger than the dynamical eccentricities derived from N-body models \citep[e.g.][]{Masuda2024}, which give $e_b = 0.0162\pm0.0038$ and $e_d = 0.0076\pm0.0058$ for the Outside 2:1 solution. The median $e_{\mathrm{min}}$ is larger than the dynamical estimate by a factor of $\sim3.4$ for b and $\sim4.4$ for d. When the two sources of uncertainty are added in quadrature, the discrepancy corresponds to $2.1\sigma$ for \exoplanet{Kepler-51}{b} and $1.1\sigma$ for \exoplanet{Kepler-51}{d}: for b, the photometric requirement is somewhat at odds with, though not inconsistent with, the dynamical limit, whereas for d the two constraints are mutually compatible.

Nevertheless, the eccentricities inferred from the TTV analysis are subject to uncertainties; so, there is a non-negligible probability that some of these eccentricities yield values of $\Psi$ that are consistent with the \photofit posterior distribution. To quantitatively evaluate this probability, we carried out a Monte Carlo calculation of the expected values of the photoeccentric bias, defined as \citepalias{Kipping2014}:

\begin{equation}
\Psi^\mathrm{PE} \equiv \frac{(1+e \sin \omega)^3}{\left(1-e^2\right)^{3 / 2}} \, .
\end{equation}

\begin{figure}[t]
\centering
\includegraphics[width=0.8\columnwidth]{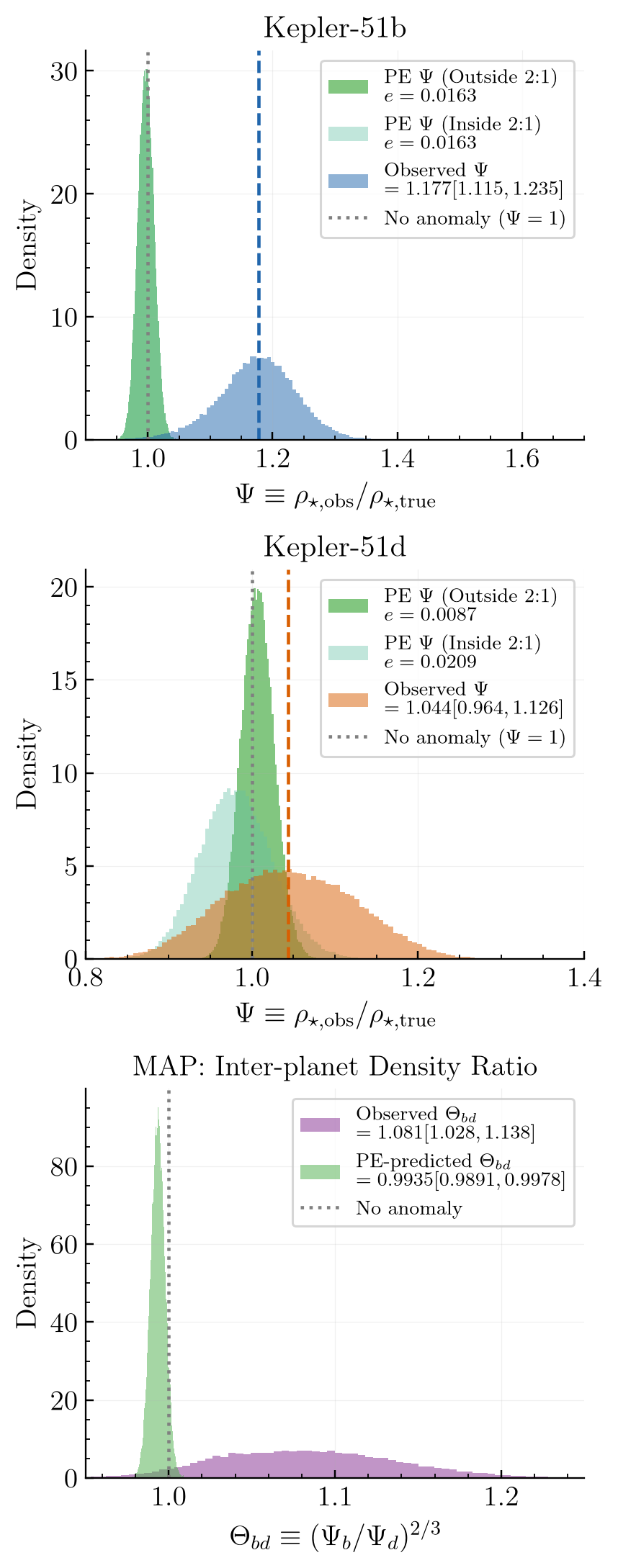}
\caption{Top panels: Forward prediction of the photoeccentric bias $\Psi^\mathrm{PE}(e,\omega)$ versus the observed density ratio $\Psi$, using dynamical posteriors from \citet{Masuda2024}. Bottom panel: The Multibody Asterodensity Profiling (MAP) analysis which compares the expected inter-planet density ratio $\Theta_{bd}=(\Psi_b/\Psi_d)^{2/3}$ to the observed one.}
\label{fig:pe_anomalies}
\end{figure}

In \autoref{fig:pe_anomalies}, we present the comparison between the distribution of predicted bias provided the uncertainties in $e$ and $e\sin\omega$ from \citet{Masuda2014} and the measured density ratio for both planets.

As is evident from the predicted PE bias distributions for both planets, the overlap with the observed density ratios is partial. To formally quantify this, we compute a $p$-value, defined as\footnote{We intentionally refer to this quantity as a $p$-value. It is a tail probability calculated under the PE hypothesis; it is not the probability that the PE hypothesis is false, and it should not be interpreted that way.} the fraction of dynamically allowed configurations that produce a synthetic density ratio $\Psi^{\mathrm{PE}}$ equal to or greater than the observed ratio $\Psi$. By marginalizing over the dynamical constraints on eccentricity and assuming a uniform distribution for the argument of periastron $\omega$, we obtain $p \approx 0.007$ for \exoplanet{Kepler-51}{b} and $p \approx 0.33$ for \exoplanet{Kepler-51}{d}. Furthermore, even under the most conservative and optimistic assumption of perfect orbital alignment ($\omega = \pi/2$, yielding the maximum possible PE bias for a given eccentricity), these values rise to $p \approx 0.033$ and $p \approx 0.42$, respectively. For \exoplanet{Kepler-51}{b} this disfavours unmodelled eccentricity as the sole driver of the offset, though we would not describe a $p$-value of this size as excluding it. For \exoplanet{Kepler-51}{d} a $p$-value of $0.33$ carries no evidential weight at all, and the photoeccentric effect remains an entirely adequate account of that planet's much smaller offset.

A final concern remains before we can fully trust the previous result. The analysis relies on the assumption that the true density is known precisely. To test for any bias arising from the adopted reference density $\rho_{\star,\mathrm{true}}$, we employ the Multibody Asterodensity Profiling (MAP) framework \citep{kipping2012novel}. MAP exploits the requirement that both planets orbit the same host star. The ratio of their transit-derived densities yields the observable quantity $\Theta_{bd} \equiv (\Psi_b/\Psi_d)^{2/3}$.
Using a first-order expansion of the photoeccentric effect, \citet{kipping2012novel} shows that a deviation of $\Theta_{bd}$ from unity imposes an absolute lower limit on the combined eccentricity of the pair:
\begin{equation}
    (e_b + e_d) \geq \frac{|\Theta_{bd} - 1|}{2} \equiv (e_b + e_d)_{\mathrm{min}}.
\end{equation}

Our analysis yields an observed ratio $\Psi_b/\Psi_d = 1.125_{-0.082}^{+0.090}$, and hence $\Theta_{bd} = 1.081^{+0.057}_{-0.053}$, which sits $1.5\sigma$ from unity with $P(\Theta_{bd} \le 1) = 5\%$. The corresponding MAP-inferred lower bound is $(e_b + e_d)_{\mathrm{min}} \approx 0.041$ $^{+0.029}_{-0.025}$. However, the sum of the dynamical eccentricities from \citet{Masuda2024} is only $(e_b + e_d) \approx 0.024$. This dynamical sum falls below the median MAP estimate but remains inside its 68\% credible interval, and $27\%$ of our posterior samples for $(e_b+e_d)_{\mathrm{min}}$ are consistent with it. The MAP test therefore does \emph{not} on its own rule out the photoeccentric effect; it yields a verdict of similarly modest significance as the single-planet analysis above. Its utility here is that it does so without relying on $\rho_{\star,\mathrm{true}}$: $\Theta_{bd}$ is constructed solely from the two transit-inferred densities, so this test is unaffected by the adopted stellar characterization and produces identical results whether one uses \citet{Masuda2024} or \citet{Berger2023}. In light of the strong dependence of the single-planet findings on that choice, we consider the MAP-based metric to be the more robust of the two.

\section{Geometric Modeling of Ringed Planet Transits} 
\label{sec:geom_model}

With the photometric offsets in hand, we now introduce the geometric forward model that maps a candidate ring configuration onto the transit observables used above.
Our inference scheme relies on a deterministic, geometry-based forward model that connects a specified ring configuration to the associated effective transit observables. While we follow that relationship via the open-source \texttt{Python} package \texttt{exorings}\footnote{The original version is available at \url{https://github.com/facom/exorings}, and an updated release is maintained in this work's repository at \url{https://github.com/seap-udea/aPRe}.} \citepalias{Zuluaga2015}, this work represents an improved and expanded version of such analytical model on planetary ring transits.

The orbital period $P$, the impact parameter $b$, and the true stellar density $\rho_{\star,\mathrm{true}}$ are three key input parameters in the particular forward-model implementation adopted in this work. From these quantities, the geometric scaled semi-major axis is obtained via Kepler’s third law (\citealt{Seager2003}, hereafter \citetalias{Seager2003}):

\begin{equation}
\left(\frac{a}{R_\star}\right)_\mathrm{geo} = \left(\frac{G\,\rho_{\star,\mathrm{true}}}{3\pi}P^2\right)^{1/3}
\label{eq:aRstar_true}
\end{equation}
where $G$ is the gravitational constant. For a circular orbit ($e=0$), the orbital inclination is then given by 

\begin{equation}
\cos i_{\rm orb}=\frac{b}{(a/R_\star)_\mathrm{geo}}  
\label{eq:iorb}
\end{equation}

Two remarks are worth highlighting. First, by using the stellar density $\rho_{\star,\mathrm{true}}$ as an input parameter instead of the stellar mass $M_\star$ and radius $R_\star$, we avoid needing prior knowledge of those basic stellar properties for the analysis itself. Second, although we explicitly include $R_\star$ when referring to the scaled semi-major axis, all spatial scales are actually expressed in units of the stellar radius. As a result, the absolute value of $R_\star$ is not strictly required for the analysis. Still, to interpret the final inferred properties or express them in SI units, we must adopt values for both stellar parameters (or at least for one of them, given that the stellar density is taken to be known).

The ring system is modeled as a geometrically thin annulus extending from an inner radius $R_{\rm in}=f_i p$ to an outer radius $R_{\rm out}=f_e p$, where $p\equiv R_p/R_\star$ corresponds to the scaled planetary radius and $f_i, f_e$ are dimensionless radii in units of $p$. 

The orientation of the ring relative to the observer is described by two angles defined with respect to the plane of the sky. These angles determine the geometry of the ring’s sky-projected shape, which is modeled as a perfect ellipse with semimajor axis $A = f_e p$ and semiminor axis $B = A \cos i_R$. The orientation angles are $i_R$, the projected ring inclination (i.e., the dihedral angle between the ring plane and the sky plane, where $i_R = 90^\circ$ represents an edge-on view of the ring), and $\theta_R$, the projected \textit{tilt} angle (the angle between the projected major axis of the ring and the orbital direction, where $\theta_R = 90^\circ$ indicates that the projected major axis is perpendicular to the transit chord). 

Planetary rings are not fully opaque. To describe their opacity, we adopt a wavelength-independent normal attenuation $\alpha$\footnote{The normal attenuation $\alpha$ is related to the normal optical depth $\tau$ via $\alpha = \exp(-\tau)$. We prefer the parameterization in terms of $\alpha$ because it is strictly bounded between $0$ and $1$. This improves numerical stability during Bayesian inference and mitigates the pathological behavior of $\tau$ in the high-opacity regime ($\tau \gg 1$), where very large variations in optical depth produce only negligible changes in the forward model.}. This parameter is then used to derive a blocking factor $\beta = 1 - \alpha^{1/\cos i_R}$ \citep{Barnes2004}, which incorporates both blocking and absorption effects and scales the ring's contribution to the total occulting area according to its inclination.

Given these inputs, our forward model computes two classes of observables.  First, it evaluates the effective occulting area $A_{\rm eff}$ and the corresponding transit depth:

\begin{equation}
\delta^\mathrm{geo}=\frac{A_{\rm eff}}{\pi R_\star^2}.
\end{equation}

In practice, $A_{\rm eff}$ is written as the area of the planetary disk plus the effective projected area of the annulus. The latter is computed using closed-form expressions for the effective projected areas of the outer and inner ring edges as functions of $(f_i,f_e, i_R, \alpha)$, including the edge-on limit where only a fraction of the annulus projects outside the planetary disk (Eq. 2 in \citetalias{Zuluaga2015}). Second, it determines the locations of the contact points along the transit chord. To illustrate the transit geometry with the exact definitions of the contact positions and transit durations, please refer to \autoref{fig:analytical_model}.

The planet-only contacts $x_{p,1\ldots4}$ are obtained from the intersection of the planetary disk with the stellar limb at impact parameter $b$. For the ring contacts $x_{R,1\ldots4}$, rather than adopting the approximate analytical intersection from \citetalias{Zuluaga2015} that breaks down at certain high inclinations, we employ a robust, exact-like analytical solution based on the Support Function of a convex body (see \autoref{sec:appendix_support_function}). The four ring-contact positions along the transit chord are given by:

\begin{eqnarray}
    x_{R,1} &\approx& -\sqrt{(1 + h_L)^2 - b^2} \nonumber \\
    x_{R,2} &\approx& -\sqrt{\max(0, [1 - h_L]^2 - b^2)} \label{eq:contact_x} \\
    x_{R,3} &\approx& +\sqrt{\max(0, [1 - h_R]^2 - b^2)} \nonumber \\
    x_{R,4} &\approx& +\sqrt{(1 + h_R)^2 - b^2}, \nonumber
\end{eqnarray}
where $h_L$ and $h_R$ are the effective directional radii of the projected ring ellipse at ingress and egress, respectively:
\begin{eqnarray}
    h_L^2 &=& A^2 (-x_0 \cos\theta_R + b \sin\theta_R)^2 \nonumber \\
          & & +\, B^2 (b \cos\theta_R + x_0 \sin\theta_R)^2 \\
    h_R^2 &=& A^2 (x_0 \cos\theta_R + b \sin\theta_R)^2 \nonumber \\
          & & +\, B^2 (b \cos\theta_R - x_0 \sin\theta_R)^2 \nonumber
\end{eqnarray}
(see equations \ref{eq:h_L} and \ref{eq:h_R} in the Appendix). These formulae are key to the methodological implementation used here: as they are fully analytical, we can perform computationally intensive numerical procedures without incurring prohibitively long computation times.

The final contact positions are then taken as $x_i=\min[x_{p,i},x_{R,i}]$, and converted into the ringed transit durations via
\begin{equation}
T_{14}^\mathrm{geo} = \frac{P}{2\pi}\arcsin\!\left[\frac{x_{4}-x_{1}}{(a/R_\star)_\mathrm{geo}\sin i_{\rm orb}}\right]
\label{eq:T14_geo}
\end{equation}
\begin{equation}
T_{23}^\mathrm{geo} = \frac{P}{2\pi}\arcsin\!\left[\frac{x_{3}-x_{2}}{(a/R_\star)_\mathrm{geo}\sin i_{\rm orb}}\right].
\label{eq:T23_geo}
\end{equation}

This mathematical procedure specifies the forward mapping from the ringed-planet parameters $(p, f_i, f_e, i_R, \theta_R, \alpha)$ to the predicted effective transit observables $(\delta^\mathrm{geo}, T_{14}^\mathrm{geo}, T_{23}^\mathrm{geo})$, under the assumption that the stellar and orbital input parameters $P$, $b$, and $\rho_{\star,\mathrm{true}}$ are independently provided.

\section{Asterodensity Profiling}
\label{sec:asterodensity_profiling}

The geometric model described in \autoref{sec:geom_model} provides the relationship between the ring parameters and the transit observables. We now introduce the asterodensity profiling framework, which converts these observables into a stellar density, along with the \PhotoRing\ effect that we invert in the remainder of this work.

Assuming a spherical, opaque planet with a mass $M_\mathrm{ p}\ll M_\star$, in a circular orbit and transiting a spherical, unblended star of radius $R_\star$ with negligible limb darkening at the maximum transit depth position, \citetalias{Seager2003} showed a mean stellar density, $\rho_{\star,\mathrm{obs}}$, can be inferred from the transit observables by inverting~\autoref{eq:aRstar_true}:

\begin{equation}
\rho_{\star,\mathrm{obs}} = \frac{3\pi}{GP^2} \left(\frac{a}{R_\star}\right)^3_{\mathrm{obs}}.
\label{eq:rho_obs}
\end{equation}

The observed scaled semi-major axis $(a/R_\star)_\mathrm{obs}$ should not be mistaken for $(a/R_\star)_\mathrm{geo}$ that is computed in the forward model from the orbital period and the true stellar density (\autoref{eq:aRstar_true}). \citetalias{Seager2003} demonstrated that $(a/R_\star)_\mathrm{obs}$ can be expressed in terms of the total $T_{14}$ and full-transit duration $T_{23}$, via their Equation 8\footnote{We distinguish here between the {\em measured} transit parameters $T_{14}$ and $T_{23}$ and the {\em calculated} values $T_{14}^\mathrm{geo}$ and $T_{23}^\mathrm{geo}$ produced by the forward model}:

\begin{equation}
\left(\frac{a}{R_\star}\right)^2_{\mathrm{obs}} = \frac{F_{(+)}^2 - b_{\mathrm{obs}}^2 \left[1 - \sin^2\left(\pi T_{14}/P\right)\right]}{\sin^2\left(\pi T_{14}/P\right)},
\label{eq:aRstar}
\end{equation}
where $F_{(\pm)}\equiv (1 \pm \sqrt{\delta})$ is an auxiliary variable. 

This formula assumes non-grazing transits ($1<b<1-p$), a planet that is much smaller than its host star ($p \ll 1$), and transit durations that are much shorter than the orbital period ($T_{14}/P \ll 1$). These conditions are typically met in most transiting exoplanet systems \citepalias{Seager2003, Kipping2014}, including the Kepler-51 planets.

Under the same assumptions, the observed impact parameter $b_{\mathrm{obs}}$ is obtained from the transit shape via Equation 9 in \citetalias{Seager2003}:

\begin{equation}
b_{\mathrm{obs}}^2 = \frac{F_{(-)}^2 - \sin^2\left(\pi T_{23}/P\right) / \sin^2\left(\pi T_{14}/P\right)F_{(+)}^2}{1 - \sin^2\left(\pi T_{23}/P\right) / \sin^2\left(\pi T_{14}/P\right)}.
\label{eq:b_obs}
\end{equation}

It is worth noting that equations (\ref{eq:aRstar}) and (\ref{eq:b_obs}), and thus \autoref{eq:rho_obs}, are equally valid when using the geometrically computed durations $T_{14}^\mathrm{geo}$ and $T_{23}^\mathrm{geo}$ from the forward model, obtained via \autoref{eq:T14_geo} and \autoref{eq:T23_geo}. The specific context in which the formulae are employed determines which set of quantities should be used. For example, when deriving the observed semi-major axis in the photometric analysis (see \autoref{sec:derived_observables}), the time and period values are those inferred from the photometric data, namely $\delta$, $T_{14}$, and $T_{23}$. In contrast, during the inference stage we employ instead the geometric quantities $\delta^\mathrm{geo}$, $T_{14}^\mathrm{geo}$, and $T_{23}^\mathrm{geo}$ (see \autoref{sec:bayes_inference}).


\subsection{The \PhotoRing\ Effect}
\label{sec:photoring_effect}

Planetary rings introduce geometric effects that are not captured by the standard spherical-planet transit model. As shown by \citetalias{Zuluaga2015}, the projected area of a ring system can substantially increase the observed transit depth $\delta_{\mathrm{obs}}$. When rings are not included in the transit model, this larger depth $\delta_{\mathrm{obs}}$ is interpreted as being produced by a larger planetary disk, which leads to an overestimation of the planetary radius $R_p$.

Rings also modify the contact times of a transit. Because the ring system extends beyond the planetary limb, the first contact occurs earlier and the fourth contact occurs later than in the case of a ringless planet. As a result, the total transit duration $T_{14}$ increases, while the full-transit duration $T_{23}$ becomes shorter. Since these three quantities constitute the fundamental transit observables, these geometric distortions propagate through the transit fitting procedure and bias the inferred values of $(a/R_\star)_{\mathrm{obs}}$ and $b_{\mathrm{obs}}$ (i.e., equations \ref{eq:aRstar} and \ref{eq:b_obs}, respectively). Consequently, the stellar density derived from the transit, $\rho_{\star,\mathrm{obs}}$, can be significantly under or overestimated. The resulting systematic offset in the stellar density inferred from transits due to an unaccounted-for ring system was first characterized by \citetalias{Zuluaga2015} and is known as the \PhotoRing\ (PR) effect. The magnitude of this effect is expressed using a logarithmic scale that employs the same notation:

\begin{equation}
\mathrm{PR} \equiv 10\,\log_{10}\!\Psi.
\label{eq:PR_def}
\end{equation}

\begin{figure*}[t]
\centering
\includegraphics[width=0.7\textwidth]{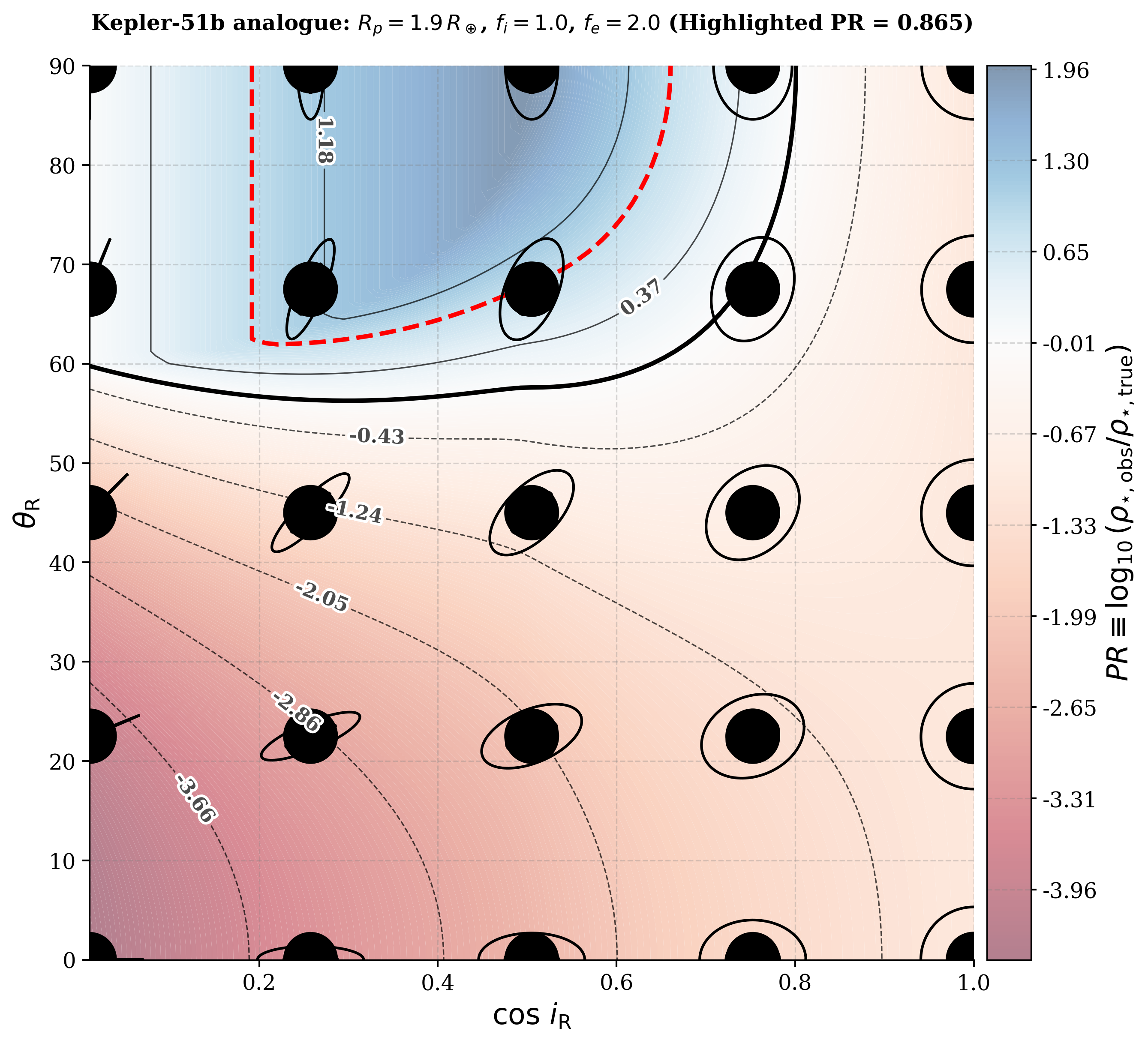}
\caption{Heat map and contour lines of the \PhotoRing\ (PR) logarithmic scale as a function of the effective ring orientation parameters $(i_{\rm R},\theta_{\rm R})$. PR is evaluated for an analogue of \exoplanet{Kepler-51}{b} with a bulk density equal to that of Earth. The thick black contour marks the zero level, $PR=0$, which identifies configurations that exhibit a density degeneracy (i.e., the inferred stellar density is equal to the true density despite the presence of rings). The highlighted contour correspond to the nominal observed offset for the actual planet \exoplanet{Kepler-51}{b} according to our \photofit fits (see text for details).}
\label{fig:photoring_contour}
\end{figure*}

In \autoref{fig:photoring_contour}, we present contours of the PR-effect metric across a wide range of ring orientations. We adopt planetary and stellar parameters akin to those of \exoplanet{Kepler-51}{b}, along with tentative properties for the ring system. As shown, PR effect typically causes the stellar density to be underestimated; that is, $\rho_{\star,\mathrm{obs}} < \rho_{\star,\mathrm{true}}$ or $\mathrm{PR}<0$, with only a relatively small domain where $\rho_{\star,\mathrm{obs}} > \rho_{\star,\mathrm{true}}$. Interestingly, for systems hosting ringed planets—with similar properties to those assumed here—the inferred stellar mean density can be underestimated by as much as a factor of $\approx 2.5$ ($\mathrm{PR}\approx -4$) or overestimated by up to a factor of $\approx 1.6$ ($\mathrm{PR}\approx +2$).

\subsection{The \PhotoRing\ Hypothesis}
\label{sec:photoring_hypothesis}

For Kepler-51, offsets in the stellar density could allegedly result from scenarios (ii) or (iii) of \autoref{sec:anomalous_stellar_densities}, but having found scenario (i) to be disfavoured for \exoplanet{Kepler-51}{b} and essentially unconstrained for \exoplanet{Kepler-51}{d} in \autoref{sec:photoeccentric_effect}, we now turn our attention to scenario (iv). 

In the remainder of this work we focus on this ringed-planet scenario—not because we consider it the most likely cause of the observed density discrepancies, but to evaluate whether a ringed configuration can bring the transit observables into agreement with a self-consistent stellar density, and to establish which ring configurations the data can already exclude. We also emphasize that this is the first study in the literature to perform a full inference using the PR effect. Therefore, even if other explanations for the density offsets are also plausible---or indeed if no effect is present at all---, this scenario offers a general methodology and establishes computational tools that can be applied to the light-curve analysis of rings in other planetary systems. 

Before proceeding, it is helpful to state the converse result, which is on more secure footing than any affirmative claim we can make here. Specifically, \emph{because} the inferred density offsets are small, any ring system around these planets must generate an equally small PR signature. The strongly negative-PR region of \autoref{fig:photoring_contour}—corresponding to extended, optically thick rings viewed at low projected tilt, and encompassing most of the allowed ring parameter space—predicts $\rho_{\star,\mathrm{obs}}$ values far below $\rho_{\star,\mathrm{true}}$ and is ruled out by our posteriors at $99\%$ credibility for \exoplanet{Kepler-51}{b} and $70\%$ for \exoplanet{Kepler-51}{d}. Consequently, a wide range of possible ring configurations is excluded using archival \textit{Kepler} photometry alone, without the need for dedicated follow-up observations.

If the density offsets are caused by the presence of rings, a notable feature of the specific case under consideration is that the sign of the measured offset in all our analyses falls in the positive-PR regime (i.e., $\rho_{\star,\mathrm{obs}} > \rho_{\star,\mathrm{true}}$), which is less frequently encountered than the negative-PR regime  (see \autoref{fig:photoring_contour}). This does not rule out the ring hypothesis—since physically plausible configurations include both zero and positive offsets—but it indicates that for rings to drive the observed offset they must fall into a relatively constrained region of the parameter space. In particular, this would require the rings to have relatively large tilt angles, $\theta_R \gtrsim 60^\circ$, with respect to the orbital plane. Such large tilts may occur when the ring inclination relative to the orbital plane is also substantial. The consequences of these configurations will be examined in \autoref{sec:discussion}. 

\section{Bayesian Inference: From Observables to Ring Parameters}
\label{sec:bayes_inference}

The forward model introduced in \autoref{sec:geom_model} specifies a deterministic relationship between the ring configuration parameters $\boldsymbol{\theta} = (\fullparameters)$—given a set of nuisance parameters $\boldsymbol{\eta} = (\fullnuisance)$—and the resulting effective transit observables: 

\begin{equation}
    \boldsymbol{\theta} \mapsto \mathbf{y}(\boldsymbol{\theta},\boldsymbol{\eta}) \equiv (\predictables).
    \label{eq:effective_observables}
\end{equation}

Our aim is to identify which ring configurations are compatible with the constraints set by the observed transits, while rigorously accounting for both observational and stellar uncertainties in the Kepler-51 system. To this end, we must specify the \textit{probability distribution function} (PDF) of the predicted parameters $(\predictables)$, conditioned on the available photometric data. 

\subsection{Likelihood}
\label{sec:kde_likelihood}

The most natural choice for the PDF of the transit observables is the joint posterior distribution of the photometric observables ($\fullkdeobservables$). You may observe that $b_\mathrm{obs}$ no longer appears in this list, as it also serves as a nuisance parameter. Because the photometric analysis does not yield a continuous distribution, we construct a non-parametric approximation to this function using a multivariate Gaussian kernel density estimator (KDE; \citealt{Scott1992}). To do so, we select a statistically representative subsample of the \photofit-derived samples of the observable vector. We denote this subsample by $\{\mathbf{y}^\mathrm{ph}_m\}_{m=1}^{N}$, where $\mathbf{y}_m$ corresponds to the mth joint realization of the chosen observables. These samples therefore provide a discrete characterization of the joint posterior distribution of $\mathbf{y}$ implied by the \photofit, which we approximate with the following Gaussian KDE: 

\begin{equation}
\widehat{p}(\mathbf{y}) = \frac{1}{N}\sum_{m=1}^{N}\,\mathcal{N}\!\left(\mathbf{y}\,\middle|\,\mathbf{y}^\mathrm{ph}_m, \mathbf{H}\right),
\label{eq:kde_def}
\end{equation}
where $\mathbf{H}=h^2\mathbf{\Sigma}$ is the KDE bandwidth matrix, constructed from the sample covariance matrix $\mathbf{\Sigma}$ and a bandwidth scaling factor $h$, which determines the degree of smoothing and the correlations of the kernel in the transit observable space. We adopt the standard automated bandwidth selection implemented in the KDE routine, following Scott's or Silverman's rule \citep{Scott1992, Silverman1986}. 

The resulting continuous function yields an estimate of the probability density across the region of transit observables preferred by the photometric analysis. For any candidate ring configuration $\boldsymbol{\theta}_k$, the forward model generates an associated observable vector, $\mathbf{y}(\boldsymbol{\theta}_k,\boldsymbol{\eta})$ (for the treatment of the nuisance parameters see \autoref{sec:bayes_nuisance} below). We then assess the agreement between this prediction and the photometric posterior by evaluating the KDE density at that point and adopting this value as the likelihood. In this way, correlated uncertainties among the observables are naturally accounted for, without depending on independent point estimates or imposing a specific parametric form for the likelihood:

\begin{equation}
\mathcal{L}(\boldsymbol{\theta}_k) \propto \widehat{p}\!\left(\mathbf{y}=\mathbf{y}(\boldsymbol{\theta}_k,\boldsymbol{\eta}\right).
\label{eq:kde_like}
\end{equation}

The retrieval process consists of sampling the posterior distribution of ring parameters, $\boldsymbol{\theta}$, exploring the region of ring-parameter space whose predicted observables are most consistent with the distribution inferred from the photometric analysis. 

\subsection{Treatment of Nuisance Parameters}
\label{sec:bayes_nuisance}

The orbital period $P$, impact parameter $b$, and true stellar density $\rho_{\star,\mathrm{true}}$ determine the transit chord geometry and orbital scale via the inferred values of $(a/R_\star)_\mathrm{geo}$ and $i_{\rm orb}$ (\autoref{eq:aRstar_true} and \autoref{eq:iorb}). They consequently influence how a given ring configuration translates into the effective transit observables. Because these parameters are not intrinsic to the ring geometry and are instead affected by measurement uncertainties, we treat them as nuisance parameters. Across our retrieval suite, we adopt several different handling strategies for them, ensuring that their uncertainties are propagated into the derived ring properties. 

In all setups, $P$ is fixed and set to the literature value, following the values reported by \citet{Masuda2024}. Depending on the specific {\bf\em run configuration}, $b$ and/or $\rho_{\star,\mathrm{true}}$ are either kept at fiducial values, explicitly sampled, or effectively marginalized over within the likelihood. When these parameters are sampled, we use a truncated normal prior for the impact parameter, $b\sim\mathrm{TruncNorm}(\mu_b,\sigma_b;[0,1])$, centered, again, on the ringless-fit value reported by \citet{Masuda2024}, and a one-dimensional KDE prior for $\rho_{\star,\mathrm{true}}$ based on the isochrone-informed stellar-density posterior from \citet{Berger2023}. 

In the most basic setup, when $b$ and/or $\rho_{\star,\mathrm{true}}$ are not sampled, we fix them at their mean values. As a robustness check, we additionally use a pseudo-marginal likelihood in which these parameters are integrated into the likelihood via Monte Carlo averaging, 

\begin{equation}
\mathcal{L}(\boldsymbol{\theta_k}) \approx \frac{1}{M}\sum_{j=1}^{M}\widehat{p}\!\left(\mathbf{y}=\mathbf{y}(\boldsymbol{\theta_k},\boldsymbol{\eta}^{(j)})\right),
\label{eq:pseudo_marg}
\end{equation}
where $\boldsymbol{\theta}$ represents the remaining ring parameters and the random realizations $\boldsymbol{\eta}^{(j)}=(P, \rho_{\star,\mathrm{true}}^{(j)},b^{(j)})$ are sampled from their prior distributions. This method incorporates the external uncertainties in $b$ and $\rho_{\star,\mathrm{true}}$ while retaining the original prior information, without embedding these parameters directly into the sampling chain. This corresponds to the proper Bayesian treatment when such parameters are external constraints that should not be updated by the transit data. 

\subsection{Parameter Space and Priors}
\label{sec:bayes_priors}

In all cases, the retrieval scheme surveys the parameter space $(\parameters)$, keeping the inner ring radius fixed at $f_i=1$ since the current photometric precision is insufficient to resolve structures lying inside the ring’s outer edge. The planetary radius ratio $p$ is either fixed or allowed to vary freely, allowing the possibility that a portion of the measured transit depth is contributed by the rings themselves. This would imply a smaller planetary radius and could partially account for the unusually low bulk densities inferred for these super-puff planets when no rings are assumed. In the fixed-radius runs, we set $p=p_{\min}$, corresponding to the limiting–density configuration for the planet (see below). 

For the planetary and ring parameters, we adopt priors within physically justified intervals. For the ring's outer radius, we adopt a uniform prior $f_e\sim\mathcal{U}(1,f_{e,\max})$. The upper bound $f_{e,\max}=10$ follows the stability limit for prograde rings, which can reach roughly $\sim0.5$ times the planetary Hill radius \citep{Domingos2006}. For the ring inclination, we assume an isotropic prior, $p(i_R)\propto\sin i_R$ for $i_R\in[0^\circ,90^\circ]$, and we adopt a uniform prior for the projected tilt, $\theta_R\sim\mathcal{U}(0^\circ,90^\circ)$, thus spanning all possible ring orientations. 

For the planetary radius, we assume a uniform prior on $p$ limited by physically motivated bounds, $p_{\min} \le p \le p_{\mathrm{obs}}$. Here, $p_{\mathrm{obs}}$ is the transit-derived radius under a ringless model (as reported by \citealt{Masuda2024}), and $p_{\min}$ is a lower bound set by requiring that the bulk density does not exceed that of a rocky, Earth-like composition, such that $p_{\min} = (R_\oplus/R_\star)\,(M_p/M_\oplus)^{1/3}$. We adopt the so-called {\em Outside 2:1} solution of \citet{Masuda2024} (their Table~6), taking $M_{p} = 6.9\,M_\oplus$ for both \exoplanet{Kepler-51}{b} and \exoplanet{Kepler-51}{d} (the reported mass ratios $m$). With $R_\star = 0.869\,R_\odot$ \citep{Berger2023}, this yields $p_{\min} = 0.0201$ for both planets ($\sim1.90\,R_\oplus$).

When we allow normal attenuation ($\alpha$) to vary freely, we impose a uniform prior $\alpha \sim \mathcal{U}(0, 1)$. For exploratory analyses, we instead fix $\alpha=0.368$, which represents a ring system with moderate opacity and is similar to the optical depths observed in portions of Saturn’s main rings \citep{Xystouris2023}. This choice prevents the introduction of an extra parameter partially degenerate with ring size and orientation, thereby lowering the dimensionality and computational expense of the retrieval while still providing a representative reference ring model. For completeness and to facilitate reproducibility, \autoref{tab:priors} reports the priors adopted in our inference procedure. 

\begin{table}[t]
\centering
\footnotesize
\caption{Prior distributions used in the ring retrieval. The planetary mass and stellar radius employed to calculate $p_\mathrm{min}$ are taken from \citet{Masuda2024} and \citet{Berger2023}, respectively.}
\label{tab:priors}
\setlength{\tabcolsep}{3pt} 
\renewcommand{\arraystretch}{1.3}
\begin{tabular}{lll}
\hline
\hline
Parameter & Prior & Range \\
\hline
\multicolumn{3}{l}{\textit{Ring geometry}} \\
$f_e$ & $\mathcal{U}(1, f_{e,\mathrm{max}})$ & $[1, 10]\,R_p$ \\
$f_i$ & Fixed & $f_i = 1$ \\
$i_R$ & $p(i_R)\propto\sin i_R$ & $[0^\circ, 90^\circ]$ \\
$\theta_R$ & $\mathcal{U}(0^\circ, 90^\circ)$ & $[0^\circ, 90^\circ]$ \\
$\alpha^{\dagger}$ & $\mathcal{U}(0, 1)$ & $[0, 1]$ \\
\hline
\multicolumn{3}{l}{\textit{Planetary radius}$^\dagger$} \\
$p$ & $\mathcal{U}(p_\mathrm{min}, p_\mathrm{obs})$ & $[0.0201, 0.0722]$ (b) \\
& & $[0.0201, 0.0986]$ (d) \\
\hline
\multicolumn{3}{l}{\textit{Nuisance parameters}$^\dagger$\footnote{Parameters indicated with a $\dagger$ are kept fixed in the baseline configurations and are only varied in the fully unconstrained model.
}} \\
$\rho_{\star,\mathrm{true}}$ & \citet{Berger2023} & $1.896^{+0.218}_{-0.246}$ \\
 $b$ & \citet{Masuda2024}  & $0.074\pm{0.072}$ (b)\\
& & $0.003\pm0.095$ (d) \\
\hline
\multicolumn{3}{l}{\textit{Fixed orbital parameters}} \\
$P$ & Fixed \citep{Masuda2024} & $45.154\,\mathrm{b}$ \\
& & $130.186\,\mathrm{d}$ \\
\hline
\hline
\end{tabular}
\end{table}

\subsection{Posterior sampling}
\label{sec:bayes_sampling}

We investigate the posterior distribution of the ring parameters using complementary Bayesian sampling techniques. Our main inference framework relies on nested sampling implemented via the \texttt{dynesty} Python package \citep{Speagle2020}, which yields both posterior samples and estimates of the Bayesian evidence ($\mathcal{Z}$). The Bayesian evidence, mathematically defined as the integral of the likelihood function over the entire prior volume ($\mathcal{Z} = \int \mathcal{L}(\boldsymbol{\theta}) \pi(\boldsymbol{\theta}) \mathrm{d}\boldsymbol{\theta}$), provides a natural metric for model comparison that intrinsically penalizes unnecessarily complex models, often referred to as the Bayesian Occam's razor. This method is particularly well suited to handling the complex likelihood structure resulting from the KDE-based observational constraints. 

We use a run configuration of $N_{\rm live} = 1200$ live points, $\Delta\ln Z = 0.01$ as stop condition, and $N_{\rm KDE}=5000$ photometric posterior samples to train the likelihood. To validate the robustness of the inferred posterior structure, we additionally analyze selected configurations using the affine-invariant ensemble MCMC sampler \texttt{emcee} \citep{ForemanMackey2013}. The agreement between both methods provides an independent consistency attention on ring parameter constraints and exploration mode.


\section{Inferred Ring Geometries} 
\label{sec:ring_results}

The problem in hand is highly dimensional: 6 dimensions correspond to the ringed planet parameters ($\fullparameters$), plus 3 nuisance parameters ($\fullnuisance$), while the likelihood-KDE is defined in an additional 4-dimensional space ($\fullkdeobservables$). With so many degrees of freedom and the added complexity of the numerical KDE, the fitting procedure may be subject to significant numerical and mathematical limitations. Our goal is to reduce this complexity while avoiding the loss of essential features of the data and the information contained therein. 

The minimal dimensionality reduction occurs when, as described in \autoref{sec:bayes_nuisance}, we treat the nuisance parameter $P$ (the orbital period) as constant and fix it to the literature value, leaving only two remaining nuisance parameters. In addition, in all configurations we impose $f_i=1$, because the current photometric precision is too low to resolve any substructure within the outer boundary of the ring. This choice further reduces the dimensionality of the parameter space to 5. Under these assumptions, the model takes as input parameters $\mathbf{\theta} =(\parameters)$, treats $\mathbf{\eta}=(\nuisance)$ as nuisance parameters, and considers $\mathbf{y}=(\fullkdeobservables)$ as observables or KDE variables. We refer to this specific choice of input+nuisance+observable parameters as the {\em full-set retrieval configuration}. 

\begin{figure*}[t]
\centering
\includegraphics[width=\textwidth,height=\textheight,keepaspectratio]{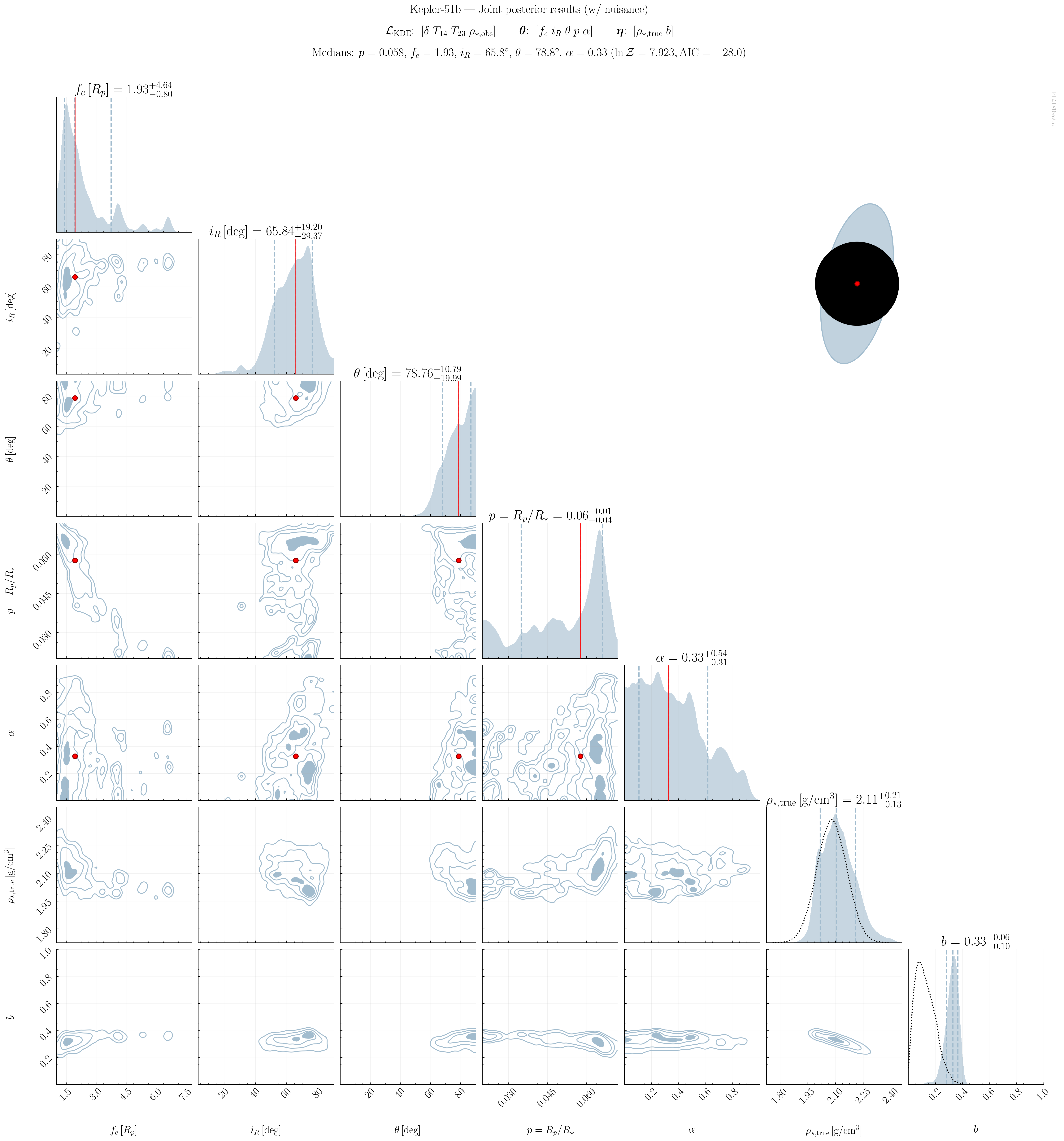}
\caption{Posterior joint distribution for the PR inference of \exoplanet{Kepler-51}{b} for a full-set retrieval setup. The corner-plot panels display 2D and 1D marginal posteriors for both the ringed planet parameters and the nuisance parameters. Together with the marginal posterior distribution of the nuisance parameters, we also show the observed PDF for both parameters (dashed lines). For visualization, we include a schematic of the ring system constructed from the median values of the marginal posteriors, indicated by red lines on the 1D panels. The dark disk denotes the planet, while the filled elliptical annulus represents the ring. Red dots on the panels indicate the ringed planet parameters used for the schematic illustration. 
}
\label{fig:full_ring_corner_b}
\end{figure*}

\begin{figure*}[t]
\centering
\includegraphics[width=\textwidth,height=\textheight,keepaspectratio]{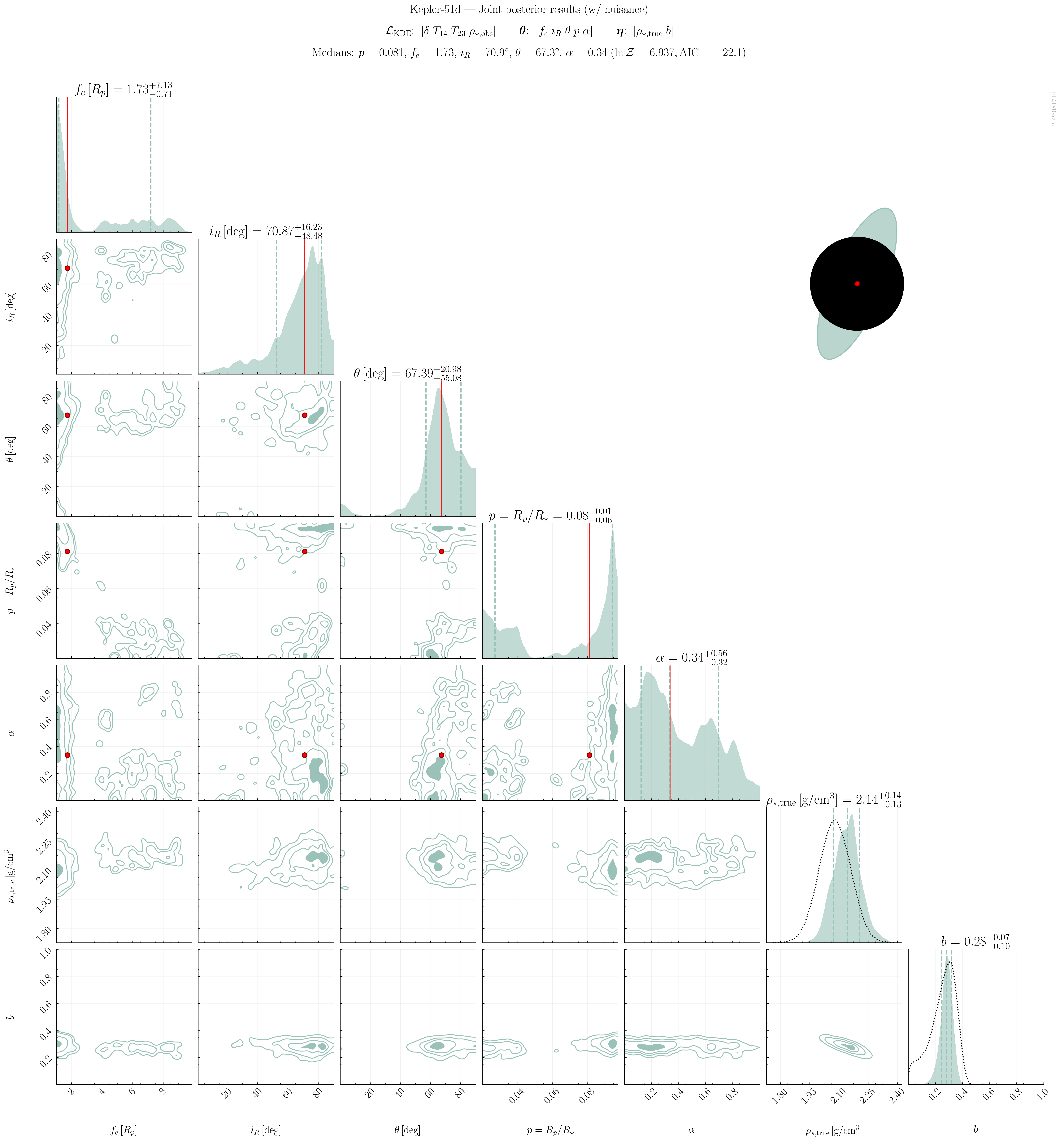}
\caption{Same as  \autoref{fig:full_ring_corner_b} but for \exoplanet{Kepler-51}{d}. 
}
\label{fig:full_ring_corner_d}
\end{figure*}

In \autoref{fig:full_ring_corner_b} and \autoref{fig:full_ring_corner_d}, we present the outcomes of fitting the data for both planets using this full-set configuration. At first glance, two notable features emerge from the schematic representation of the planet in the corner plot and the median values reported in the figure title (see also \autoref{tab:full_set_results}). First, the planetary depiction is consistent with our expectation that, since PR\,$>$\,0 for both planets, the favored ring configurations correspond to highly inclined rings located in the upper part of \autoref{fig:photoring_contour}. Second, we find that the resulting solutions yield planetary radii comparable to those inferred from the ring-free analysis: this indicates that the ringed model likewise requires inflated planets, now encircled by rings. We call these solutions ``puff ringed planets''. 

In the same figures, it is worth noting the tension between the values of the true density when it is treated as a nuisance parameter (see filled area in the marginal plot, second row from bottom to top) and the independent estimate of this same quantity (dashed curve). An analogous tension appears for the nuisance parameter $b$. Since these two parameters set the orbital scale and the transit chord, respectively, their combined freedom enables the forward model to trade off the ring-contact geometry against the transit durations, without artificially forcing the ring parameters to absorb a geometric discrepancy they cannot physically account for. Nevertheless, the sampling procedure fails to obtain solutions fully consistent with the observed data when the assumed true stellar densities are too low. Notably, this tension could, in principle, be exploited to constrain the true stellar density using this method. 

\begin{table}[t]
\centering
\footnotesize
\caption{Full-set retrieval results for \exoplanet{Kepler-51}{b} and \exoplanet{Kepler-51}{d}. The table reports the median values and the 16th and 84th percentiles for the ring geometry parameters, nuisance parameters, and selected derived physical properties. We assume a stellar radius of $0.869\,R_\odot$ \citep{Berger2023} and a planetary mass of $6.9\,M_\oplus$ for both planets \citep{Masuda2024}. For comparison, under a ringless model, the inferred planetary radii are $6.85\,R_\oplus$ (\exoplanet{Kepler-51}{b}) and $9.36\,R_\oplus$ (\exoplanet{Kepler-51}{d}), yielding anomalous densities of $0.11\,\mathrm{g\,cm^{-3}}$ and $0.038\,\mathrm{g\,cm^{-3}}$, respectively.}
\label{tab:full_set_results}
\setlength{\tabcolsep}{3pt} 
\renewcommand{\arraystretch}{1.3}
\begin{tabular*}{\columnwidth}{@{\extracolsep{\fill}}lcc}
\hline
\hline
Parameter & \exoplanet{Kepler-51}{b} & \exoplanet{Kepler-51}{d} \\
\hline
\multicolumn{3}{c}{\textit{Ring geometry}} \\
$f_e$ [$R_p$] & $1.53^{+3.11}_{-0.40}$ & $1.84^{+5.25}_{-0.56}$ \\
$i_R$ [$^\circ$] & $69.8^{+13.1}_{-11.0}$ & $63.7^{+15.0}_{-21.8}$ \\
$\theta_R$ [$^\circ$] & $73.8^{+12.7}_{-20.2}$ & $71.1^{+11.8}_{-20.2}$ \\
$p$ & $0.0664^{+0.0036}_{-0.0408}$ & $0.0721^{+0.0210}_{-0.0392}$ \\
$\alpha$ & $0.236^{+0.361}_{-0.165}$ & $0.473^{+0.325}_{-0.312}$ \\
\hline
\multicolumn{3}{l}{\textit{Nuisance parameters}} \\
$\rho_{\star,\mathrm{true}}$ [$\mathrm{g\,cm^{-3}}$] & $2.31^{+0.14}_{-0.35}$ & $2.16^{+0.20}_{-0.18}$ \\
$b$ & $0.229^{+0.157}_{-0.111}$ & $0.271^{+0.084}_{-0.147}$ \\
\midrule
\multicolumn{3}{l}{\textit{Derived parameters}} \\
$R_p$ [$R_\oplus$] & 6.30 & 6.84 \\
$\rho_p$ [$\mathrm{g\,cm^{-3}}$] & 0.152 & 0.119 \\
\hline
\hline
\end{tabular*}
\end{table}

\subsection{Predictive Posterior Checks}
\label{sec:ring_predictive}

Finding a solution to the retrieval problem is not, by itself, sufficient to ensure that the inferred model actually explains the observed transit features—most importantly if we want to ultimately account for the measured stellar density. To assess whether the nested samples produced by our retrievals can recover the marginal posterior distributions of the key observables from the photometric analysis, we conduct goodness-of-fit evaluations using posterior predictive checks (PPCs).

Specifically, for each retrieval configuration we propagate the corresponding posterior samples through the forward model presented in \autoref{sec:geom_model}. We then compare the resulting marginal posterior distributions of the transit observables with those derived directly from the photometric data (\autoref{sec:data_analysis}). For additional validation, we also consider observables that were not explicitly included in the likelihood function. The agreement for each observable is quantified using two statistics (see, e.g., \citealt{Ramdas2017}): (i) the 1-Wasserstein distance $W_1$, which measures the mean absolute displacement between the two empirical distributions; and (ii) the energy distance $E$, which measures the squared discrepancy between the two distributions. 

\begin{figure*}[t]
    \centering
    \includegraphics[width=0.65\linewidth]{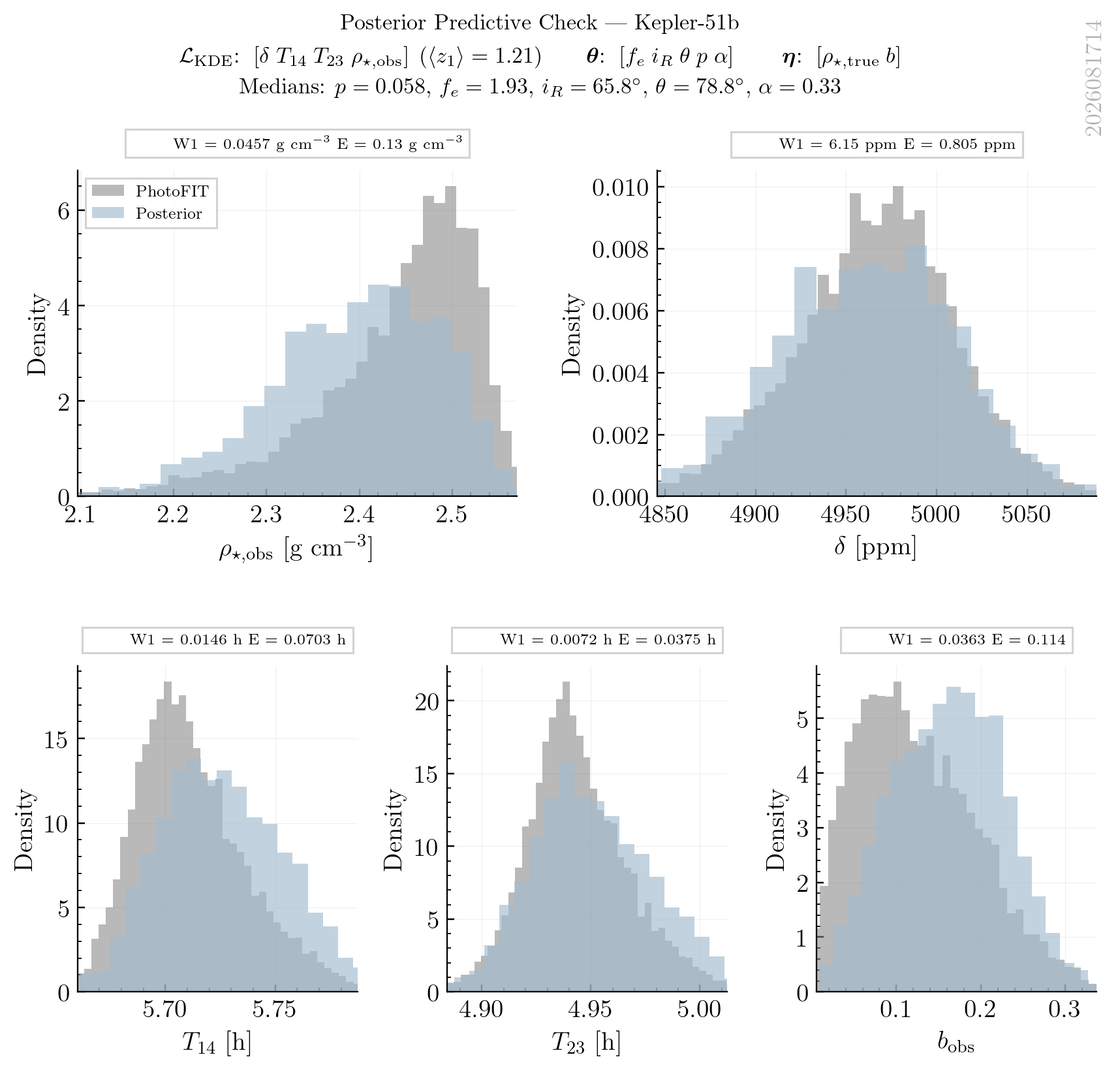}
    \includegraphics[width=0.65\linewidth]{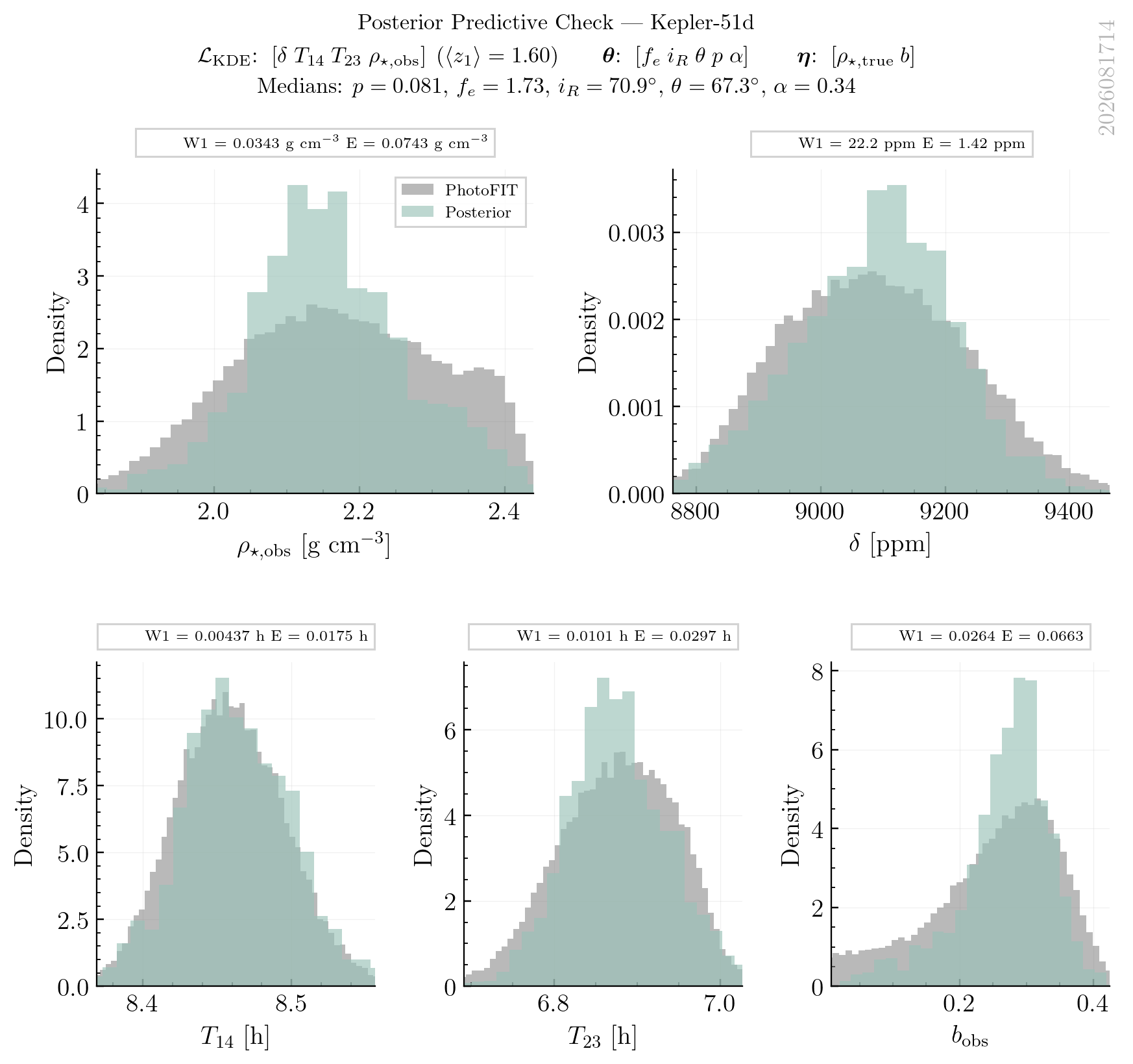}
    \caption{Posterior predictive checks (PPCs) for \exoplanet{Kepler-51}{b} and \exoplanet{Kepler-51}{d} in the full-set retrieval setup. PPCs compare the model-predicted distribution (blue) with the photometrically derived posterior (grey) for each transit observable. The reported values of $W_1$ and $E$ (see main text) quantify the self-consistent goodness-of-fit between the ringed-planet model and the observed transit observables.
    }
    \label{fig:full_ppcs_planet}
\end{figure*}

In \autoref{fig:full_ppcs_planet}, we compare the marginal posterior distributions (displayed as histograms) of the observables $(\fullkdeobservables)$ derived from the photometric analysis with those obtained from the full-set retrieval configuration, whose selected outcomes are shown in \autoref{fig:full_ring_corner_b} and \autoref{fig:full_ring_corner_d}. For both planets, the ring model reproduces the observed distributions of all four observables, a conclusion further supported by the corresponding $W_1$ and $E$ goodness-of-fit metrics. We emphasize that this concerns the adaptability of the ringed model rather than the evidence for the presence of rings: a model without rings, but with suitably tuned stellar density, can reproduce the same observables equally well. 

\subsection{Grid search}
\label{sec:grid_search}

After inspecting the solution obtained from the full-set retrieval, several characteristics of the parameter space in which the problem is defined become apparent. First, for both planets we identify regions of parameter space where one or two values of the planetary radius stand out with higher probabilities than the rest (see third row of \autoref{fig:full_ring_corner_b} and \autoref{fig:full_ring_corner_d}, top to bottom). 

For \exoplanet{Kepler-51}{b}, we find a radius close to the minimum allowed value (first peak in the marginal distribution of the third row of \autoref{fig:full_ring_corner_b}). When compared with the marginal posterior in the $p$–$f_e$ plane (first column, third row), this corresponds to a large ring with $f_e \approx 4$ (second peak in the upper marginal distribution of $f_e$, first column) and a relatively modest opacity (maximum in the $\alpha$–$p$ panel also in the fourth row). The second solution corresponds to a planet whose radius is nearly identical to that obtained from the ringless model (second peak in the marginal distribution of the third row). As anticipated, in this configuration the ring radius can be very small (first peak in the upper marginal distribution of $f_e$, first column), and the opacity is essentially unconstrained (islands in the $\alpha$–$p$ marginal of the fourth row). An analogous analysis can be performed for \exoplanet{Kepler-51}{d}, based on an examination of the panels in \autoref{fig:full_ring_corner_d} and the insights obtained from the marginal distributions.

This may indicate that our search should better focus on limited regions of the parameter space where promising solutions tend to cluster, rather than taking the median values of the posterior distributions as the best fit. To investigate these specific regions, we performed a grid search. For this exploration, we performed retrievals where the planetary radius $p$ and $\alpha$ were fixed on a grid composed of 5 values of $p$, evenly distributed between the minimum value $p_\mathrm{min}$ and the observed value $p_\mathrm{obs}$ (excluding both endpoints), and three values of $\alpha$ (0.2, 0.5, 0.8). The selection of the grid resolution and its specific values was rather arbitrary, and a more exhaustive investigation could be undertaken. Our primary aim was to probe these regions to identify potential local likelihood maxima.

In \autoref{tab:grid_golden} we present the best retrieval results in the grid search for both planets. The retrievals shown are those that fulfill a set of conditions evaluating the quality of the fit, including excellent posterior predictive checks (with normalized deviations $z_1 \lesssim 2$ and $z_\mathrm{crit} \lesssim 2$ in the transit observables), a tight agreement with the independently derived true stellar density ($\Delta \rho_{\star,\mathrm{true}} \lesssim 0.3\,\mathrm{g\,cm^{-3}}$), strongly peaked posteriors for the ring inclination, and a lower limit on the ring outer radius indicating a physically distinct structure ($f_e > 1.1$ at the 16th percentile). We refer to this set of optimal solutions as the ``Golden Sample''.

\begin{table*}[t]
\centering
\footnotesize
\caption{Golden Sample retrievals from the radius-alpha grid search for \exoplanet{Kepler-51}{b} and \exoplanet{Kepler-51}{d}.}
\label{tab:grid_golden}
\setlength{\tabcolsep}{4pt}
\begin{tabular*}{\textwidth}{@{\extracolsep{\fill}} c @{\hspace{0.3em}} c @{\hspace{0.3em}} c @{\hspace{0.3em}} c @{\hspace{0.3em}} c @{\hspace{0.3em}} c @{\hspace{0.3em}} c @{\hspace{0.3em}} c @{\hspace{0.3em}} c @{\hspace{0.3em}} c @{\hspace{0.3em}} c @{}}
\toprule
& \multicolumn{7}{c}{Input selection} & Other & Metrics & Joint pdf \\
\cmidrule(lr){2-8} \cmidrule(lr){9-9} \cmidrule(lr){10-10} \cmidrule(lr){11-11}
Planet & $p\;[R_\star]\;(R_\oplus)$ & $f_e\;[R_p]$ & $i_R\;[^\circ]$ & $\theta_R\;[^\circ]$ & $\alpha$ & $\rho_{\star,\mathrm{true}}\:[\mathrm{g\,cm^{-3}}]$ & $b$ & $\rho_p\:[\mathrm{g\,cm^{-3}}]$ & $\ln \mathcal{Z}$ & Plot link \\
\midrule
\multirow{6}{*}{\rotatebox{90}{\exoplanet{Kepler-51}{b}}} & $0.0462$\;(4.38) & $3.52^{+1.23}_{-0.98}$ & $80.6^{+4.8}_{-11.5}$ & $80.0^{+5.7}_{-9.8}$ & $0.20$ & $2.15^{+0.07}_{-0.09}$ & $0.31^{+0.04}_{-0.04}$ & 0.452 & 8.17 & \href{https://github.com/seap-udea/PRisma/blob/master/pipeline/kepler_51/results/exorings/explore_radius_alpha_masuda/./kepler_51_b_cat1_GoldenSample_ord91829-lnZ_%2B08.17-kepler_51_b_NS_exorings_kde_delta-T14-T23-rho_obs_nlive1200_dlogz0.01_NKDE5000_seed2026_rhoFREE_bFREE_rhoMasuda_P5_A3_corner.png}{pdf} \textbar \href{https://github.com/seap-udea/PRisma/blob/master/pipeline/kepler_51/results/exorings/explore_radius_alpha_masuda/./kepler_51_b_cat1_GoldenSample_ord91829-lnZ_%2B08.17-kepler_51_b_NS_exorings_kde_delta-T14-T23-rho_obs_nlive1200_dlogz0.01_NKDE5000_seed2026_rhoFREE_bFREE_rhoMasuda_P5_A3_ppc.png}{ppc} \\
 & $0.0566$\;(5.37) & $1.72^{+0.95}_{-0.26}$ & $59.4^{+22.4}_{-18.7}$ & $76.0^{+7.7}_{-9.7}$ & $0.20$ & $2.16^{+0.08}_{-0.09}$ & $0.31^{+0.04}_{-0.04}$ & 0.245 & 8.13 & \href{https://github.com/seap-udea/PRisma/blob/master/pipeline/kepler_51/results/exorings/explore_radius_alpha_masuda/./kepler_51_b_cat1_GoldenSample_ord91874-lnZ_%2B08.13-kepler_51_b_NS_exorings_kde_delta-T14-T23-rho_obs_nlive1200_dlogz0.01_NKDE5000_seed2026_rhoFREE_bFREE_rhoMasuda_P7_A3_corner.png}{pdf} \textbar \href{https://github.com/seap-udea/PRisma/blob/master/pipeline/kepler_51/results/exorings/explore_radius_alpha_masuda/./kepler_51_b_cat1_GoldenSample_ord91874-lnZ_%2B08.13-kepler_51_b_NS_exorings_kde_delta-T14-T23-rho_obs_nlive1200_dlogz0.01_NKDE5000_seed2026_rhoFREE_bFREE_rhoMasuda_P7_A3_ppc.png}{ppc} \\
 & $0.0566$\;(5.37) & $1.98^{+0.55}_{-0.23}$ & $67.4^{+13.0}_{-12.8}$ & $78.4^{+6.8}_{-6.0}$ & $0.50$ & $2.13^{+0.08}_{-0.08}$ & $0.32^{+0.04}_{-0.04}$ & 0.245 & 8.07 & \href{https://github.com/seap-udea/PRisma/blob/master/pipeline/kepler_51/results/exorings/explore_radius_alpha_masuda/./kepler_51_b_cat1_GoldenSample_ord91934-lnZ_%2B08.07-kepler_51_b_NS_exorings_kde_delta-T14-T23-rho_obs_nlive1200_dlogz0.01_NKDE5000_seed2026_rhoFREE_bFREE_rhoMasuda_P7_A2_corner.png}{pdf} \textbar \href{https://github.com/seap-udea/PRisma/blob/master/pipeline/kepler_51/results/exorings/explore_radius_alpha_masuda/./kepler_51_b_cat1_GoldenSample_ord91934-lnZ_%2B08.07-kepler_51_b_NS_exorings_kde_delta-T14-T23-rho_obs_nlive1200_dlogz0.01_NKDE5000_seed2026_rhoFREE_bFREE_rhoMasuda_P7_A2_ppc.png}{ppc} \\
 & $0.0357$\;(3.39) & $3.57^{+0.95}_{-0.45}$ & $70.1^{+8.8}_{-10.0}$ & $75.8^{+8.1}_{-6.0}$ & $0.50$ & $2.14^{+0.07}_{-0.06}$ & $0.32^{+0.03}_{-0.03}$ & 0.976 & 7.66 & \href{https://github.com/seap-udea/PRisma/blob/master/pipeline/kepler_51/results/exorings/explore_radius_alpha_masuda/./kepler_51_b_cat1_GoldenSample_ord92341-lnZ_%2B07.66-kepler_51_b_NS_exorings_kde_delta-T14-T23-rho_obs_nlive1200_dlogz0.01_NKDE5000_seed2026_rhoFREE_bFREE_rhoMasuda_P3_A2_corner.png}{pdf} \textbar \href{https://github.com/seap-udea/PRisma/blob/master/pipeline/kepler_51/results/exorings/explore_radius_alpha_masuda/./kepler_51_b_cat1_GoldenSample_ord92341-lnZ_%2B07.66-kepler_51_b_NS_exorings_kde_delta-T14-T23-rho_obs_nlive1200_dlogz0.01_NKDE5000_seed2026_rhoFREE_bFREE_rhoMasuda_P3_A2_ppc.png}{ppc} \\
 & $0.0462$\;(4.38) & $3.17^{+0.67}_{-0.71}$ & $77.5^{+4.9}_{-13.9}$ & $78.4^{+6.6}_{-7.0}$ & $0.50$ & $2.14^{+0.07}_{-0.06}$ & $0.32^{+0.03}_{-0.04}$ & 0.452 & 7.25 & \href{https://github.com/seap-udea/PRisma/blob/master/pipeline/kepler_51/results/exorings/explore_radius_alpha_masuda/./kepler_51_b_cat1_GoldenSample_ord92749-lnZ_%2B07.25-kepler_51_b_NS_exorings_kde_delta-T14-T23-rho_obs_nlive1200_dlogz0.01_NKDE5000_seed2026_rhoFREE_bFREE_rhoMasuda_P5_A2_corner.png}{pdf} \textbar \href{https://github.com/seap-udea/PRisma/blob/master/pipeline/kepler_51/results/exorings/explore_radius_alpha_masuda/./kepler_51_b_cat1_GoldenSample_ord92749-lnZ_%2B07.25-kepler_51_b_NS_exorings_kde_delta-T14-T23-rho_obs_nlive1200_dlogz0.01_NKDE5000_seed2026_rhoFREE_bFREE_rhoMasuda_P5_A2_ppc.png}{ppc} \\
 & $0.0566$\;(5.37) & $2.56^{+0.25}_{-0.16}$ & $74.3^{+6.3}_{-6.8}$ & $83.5^{+4.0}_{-4.5}$ & $0.80$ & $2.15^{+0.07}_{-0.07}$ & $0.31^{+0.03}_{-0.04}$ & 0.245 & 7.24 & \href{https://github.com/seap-udea/PRisma/blob/master/pipeline/kepler_51/results/exorings/explore_radius_alpha_masuda/./kepler_51_b_cat1_GoldenSample_ord92757-lnZ_%2B07.24-kepler_51_b_NS_exorings_kde_delta-T14-T23-rho_obs_nlive1200_dlogz0.01_NKDE5000_seed2026_rhoFREE_bFREE_rhoMasuda_P7_A1_corner.png}{pdf} \textbar \href{https://github.com/seap-udea/PRisma/blob/master/pipeline/kepler_51/results/exorings/explore_radius_alpha_masuda/./kepler_51_b_cat1_GoldenSample_ord92757-lnZ_%2B07.24-kepler_51_b_NS_exorings_kde_delta-T14-T23-rho_obs_nlive1200_dlogz0.01_NKDE5000_seed2026_rhoFREE_bFREE_rhoMasuda_P7_A1_ppc.png}{ppc} \\
\midrule
\multirow{6}{*}{\rotatebox{90}{\exoplanet{Kepler-51}{d}}} & $0.0907$\;(8.61) & $1.35^{+0.32}_{-0.11}$ & $67.5^{+15.6}_{-20.0}$ & $68.8^{+14.5}_{-14.5}$ & $0.50$ & $2.12^{+0.08}_{-0.07}$ & $0.29^{+0.04}_{-0.05}$ & 0.060 & 7.35 & \href{https://github.com/seap-udea/PRisma/blob/master/pipeline/kepler_51/results/exorings/explore_radius_alpha_masuda/./kepler_51_d_cat1_GoldenSample_ord92648-lnZ_%2B07.35-kepler_51_d_NS_exorings_kde_delta-T14-T23-rho_obs_nlive1200_dlogz0.01_NKDE5000_seed2026_rhoFREE_bFREE_rhoMasuda_P9_A2_corner.png}{pdf} \textbar \href{https://github.com/seap-udea/PRisma/blob/master/pipeline/kepler_51/results/exorings/explore_radius_alpha_masuda/./kepler_51_d_cat1_GoldenSample_ord92648-lnZ_%2B07.35-kepler_51_d_NS_exorings_kde_delta-T14-T23-rho_obs_nlive1200_dlogz0.01_NKDE5000_seed2026_rhoFREE_bFREE_rhoMasuda_P9_A2_ppc.png}{ppc} \\
 & $0.0279$\;(2.65) & $4.29^{+1.05}_{-0.27}$ & $45.8^{+19.8}_{-12.1}$ & $69.5^{+11.8}_{-12.6}$ & $0.20$ & $2.11^{+0.10}_{-0.09}$ & $0.30^{+0.05}_{-0.05}$ & 2.046 & 6.14 & \href{https://github.com/seap-udea/PRisma/blob/master/pipeline/kepler_51/results/exorings/explore_radius_alpha_masuda/./kepler_51_d_cat1_GoldenSample_ord93862-lnZ_%2B06.14-kepler_51_d_NS_exorings_kde_delta-T14-T23-rho_obs_nlive1200_dlogz0.01_NKDE5000_seed2026_rhoFREE_bFREE_rhoMasuda_P1_A3_corner.png}{pdf} \textbar \href{https://github.com/seap-udea/PRisma/blob/master/pipeline/kepler_51/results/exorings/explore_radius_alpha_masuda/./kepler_51_d_cat1_GoldenSample_ord93862-lnZ_%2B06.14-kepler_51_d_NS_exorings_kde_delta-T14-T23-rho_obs_nlive1200_dlogz0.01_NKDE5000_seed2026_rhoFREE_bFREE_rhoMasuda_P1_A3_ppc.png}{ppc} \\
 & $0.0279$\;(2.65) & $5.62^{+1.65}_{-0.36}$ & $62.0^{+14.7}_{-8.3}$ & $71.2^{+12.9}_{-5.7}$ & $0.50$ & $2.13^{+0.07}_{-0.08}$ & $0.29^{+0.04}_{-0.04}$ & 2.046 & 6.07 & \href{https://github.com/seap-udea/PRisma/blob/master/pipeline/kepler_51/results/exorings/explore_radius_alpha_masuda/./kepler_51_d_cat1_GoldenSample_ord93935-lnZ_%2B06.07-kepler_51_d_NS_exorings_kde_delta-T14-T23-rho_obs_nlive1200_dlogz0.01_NKDE5000_seed2026_rhoFREE_bFREE_rhoMasuda_P1_A2_corner.png}{pdf} \textbar \href{https://github.com/seap-udea/PRisma/blob/master/pipeline/kepler_51/results/exorings/explore_radius_alpha_masuda/./kepler_51_d_cat1_GoldenSample_ord93935-lnZ_%2B06.07-kepler_51_d_NS_exorings_kde_delta-T14-T23-rho_obs_nlive1200_dlogz0.01_NKDE5000_seed2026_rhoFREE_bFREE_rhoMasuda_P1_A2_ppc.png}{ppc} \\
 & $0.0593$\;(5.63) & $4.16^{+1.03}_{-0.54}$ & $79.8^{+5.4}_{-8.2}$ & $82.0^{+5.0}_{-5.7}$ & $0.80$ & $2.12^{+0.07}_{-0.06}$ & $0.29^{+0.03}_{-0.04}$ & 0.213 & 5.55 & \href{https://github.com/seap-udea/PRisma/blob/master/pipeline/kepler_51/results/exorings/explore_radius_alpha_masuda/./kepler_51_d_cat1_GoldenSample_ord94452-lnZ_%2B05.55-kepler_51_d_NS_exorings_kde_delta-T14-T23-rho_obs_nlive1200_dlogz0.01_NKDE5000_seed2026_rhoFREE_bFREE_rhoMasuda_P5_A1_corner.png}{pdf} \textbar \href{https://github.com/seap-udea/PRisma/blob/master/pipeline/kepler_51/results/exorings/explore_radius_alpha_masuda/./kepler_51_d_cat1_GoldenSample_ord94452-lnZ_%2B05.55-kepler_51_d_NS_exorings_kde_delta-T14-T23-rho_obs_nlive1200_dlogz0.01_NKDE5000_seed2026_rhoFREE_bFREE_rhoMasuda_P5_A1_ppc.png}{ppc} \\
 & $0.0436$\;(4.14) & $3.41^{+0.51}_{-0.15}$ & $57.7^{+11.6}_{-6.3}$ & $71.9^{+10.0}_{-7.8}$ & $0.50$ & $2.14^{+0.08}_{-0.07}$ & $0.28^{+0.04}_{-0.05}$ & 0.536 & 4.94 & \href{https://github.com/seap-udea/PRisma/blob/master/pipeline/kepler_51/results/exorings/explore_radius_alpha_masuda/./kepler_51_d_cat1_GoldenSample_ord95064-lnZ_%2B04.94-kepler_51_d_NS_exorings_kde_delta-T14-T23-rho_obs_nlive1200_dlogz0.01_NKDE5000_seed2026_rhoFREE_bFREE_rhoMasuda_P3_A2_corner.png}{pdf} \textbar \href{https://github.com/seap-udea/PRisma/blob/master/pipeline/kepler_51/results/exorings/explore_radius_alpha_masuda/./kepler_51_d_cat1_GoldenSample_ord95064-lnZ_%2B04.94-kepler_51_d_NS_exorings_kde_delta-T14-T23-rho_obs_nlive1200_dlogz0.01_NKDE5000_seed2026_rhoFREE_bFREE_rhoMasuda_P3_A2_ppc.png}{ppc} \\
 & $0.0436$\;(4.14) & $5.65^{+0.88}_{-0.46}$ & $77.6^{+4.9}_{-5.9}$ & $81.5^{+5.7}_{-5.5}$ & $0.80$ & $2.13^{+0.06}_{-0.05}$ & $0.29^{+0.02}_{-0.03}$ & 0.536 & 4.52 & \href{https://github.com/seap-udea/PRisma/blob/master/pipeline/kepler_51/results/exorings/explore_radius_alpha_masuda/./kepler_51_d_cat1_GoldenSample_ord95481-lnZ_%2B04.52-kepler_51_d_NS_exorings_kde_delta-T14-T23-rho_obs_nlive1200_dlogz0.01_NKDE5000_seed2026_rhoFREE_bFREE_rhoMasuda_P3_A1_corner.png}{pdf} \textbar \href{https://github.com/seap-udea/PRisma/blob/master/pipeline/kepler_51/results/exorings/explore_radius_alpha_masuda/./kepler_51_d_cat1_GoldenSample_ord95481-lnZ_%2B04.52-kepler_51_d_NS_exorings_kde_delta-T14-T23-rho_obs_nlive1200_dlogz0.01_NKDE5000_seed2026_rhoFREE_bFREE_rhoMasuda_P3_A1_ppc.png}{ppc} \\
\bottomrule
\end{tabular*}
\end{table*}

As shown in \autoref{tab:grid_golden}, several of the local configurations with fixed $p$ and $\alpha$ yield a higher Bayesian evidence ($\ln \mathcal{Z}$) than the corresponding full-set retrieval where these parameters vary freely. This behavior directly arises from the Occam's razor effect inherent to the Bayesian evidence that integrates the likelihood over the entire prior volume. In the highly-degenerate full-set retrieval, the sampler must explore a much larger prior space, including regions of low likelihood and penalizing the total evidence. By fixing $p$ and $\alpha$ near a local likelihood peak, the prior volume is drastically reduced in those dimensions, thus avoiding the volume penalty. Additionally, breaking the degeneracies between the ring geometry and the planetary radius improves the numerical efficiency of the Nested Sampling algorithm, allowing it to better resolve the local probability maxima.


\section{DISCUSSION AND CONCLUSIONS}
\label{sec:discussion}

The framework developed in this study was intended to evaluate whether ring configurations can be found that remain compatible with the existing Kepler-51 transit observations—taking into account the $\sim 30$\,min instrumental cadence—and, in particular, with the observed density offsets. 

Before turning to the inferred geometries, it is helpful to state clearly what the data do and do not support. Using the stellar density from \citet{Masuda2024}, the transit-inferred value is higher than the isochrone-based estimate by $2.9\sigma$ for \exoplanet{Kepler-51}{b} and by $0.6\sigma$ for \exoplanet{Kepler-51}{d}. Only the former constitutes a meaningful (though still sub-detection) preference for a positive PR offset; \exoplanet{Kepler-51}{d} alone remains compatible with no effect. We also emphasize that these significance levels depend substantially on the choice of stellar characterization—the \citet{Berger2023} posterior instead implies $2.0\sigma$ and $1.1\sigma$—so at present the stellar density, rather than the photometry, is the main limiting factor for \exoplanet{Kepler-51}{b}. Accordingly, the ring retrieval above should be interpreted in two complementary ways: both as a complete, worked example of the PR framework—here applied to real data for the first time—and as a null-result style exclusion test. The non-detection of a large density offset is itself an informative constraint, ruling out a wide swath of ring parameter space, particularly extended, optically thick, highly inclined systems that would generate the strongest photometric signals. In our view, the fact that such constraints can be drawn from a decade-old archival photometric dataset, without any dedicated space-based follow-up, is the most generally applicable result of this study.

Our Bayesian retrievals identify ring configurations that can replicate the complete set of derived transit observables for both planets, as obtained from a photometric fit with independent transit times, i.e. \photofit.. However, these solutions require both $\rho_{\star,\mathrm{true}}$ and $b$ inconsistent with their current best external constraints from \citet{Berger2023} and \citet{Masuda2024}, preventing a fully self-consistent ring solution from \textit{Kepler} photometry alone. This reflects the intrinsic degeneracy of the problem: numerical convergence toward compatible ring geometries does not guarantee physical viability, which ultimately depends on tighter independent constraints.

It is important emphasizing that our analysis is not a direct fit of a ringed-planet transit model to the \textit{Kepler} light curves. Instead, it is based on the global transit observables set $(\observables)$, which captures only part of the information contained in the full transit morphology. A direct fit to the photometry, ideally performed jointly with a dynamical TTV model, would exploit the detailed ingress and egress shape and any residual signatures to further constrain the ring geometry and break some of the remaining parameter degeneracies in our analysis.

Although we do not obtain a fully self-consistent ring solution, our preferred retrievals for both planets converge toward ring systems with variable extensions ($f_e\approx 1.3$--$6.2\,R_p$ for \exoplanet{Kepler-51}{b} and $f_e\approx 1.3$--$3.9\,R_p$ for \exoplanet{Kepler-51}{d}), moderately inclined ring planes ($i_R\approx 50^\circ$--$80^\circ$), and similarly moderate projected tilts ($\theta_R\approx 60^\circ$--$80^\circ$). Among the explored solutions, we favor this configuration over either nearly edge-on or face-on geometries, as it is also more consistent with current theoretical expectations for ring formation and long-term stability \citep[see e.g.,][]{Canup2010,Crida2019,Estrada2023,Saillenfest2023}. The ring interpretation further suggests a potential detectability bias. As illustrated by our forward model and \PhotoRing-effect map (\autoref{fig:photoring_contour}), the strongest photometric signatures arise from large, optically thick rings observed at high obliquity. The configurations preferred by our retrieval instead occupy a region of parameter space where the expected photometric signatures are comparatively weaker. If similar ring geometries are common among exoplanets, current searches would be intrinsically less sensitive to them, suggesting that at least part of the apparent lack of confirmed exoplanetary rings may arise from observational selection effects rather than from genuinely low ring occurrence rates.

The anomalous bulk densities under the ring hypothesis neither exclude alternative explanations for the inflated radii of the Kepler-51 planets nor are they mutually exclusive with them. Atmospheric inflation mechanisms such as high-altitude hazes, primordial H/He envelopes, or a young evolutionary state remain viable scenarios \citep{Libby-Roberts2020, Wang2019, Lammers2024}. Similarly, while the flat transmission spectra observed with HST and JWST are commonly attributed to aerosols that mute molecular features \citep{Libby-Roberts2025}, an optically thick ring system could produce the same effect by attenuating stellar light achromatically. Distinguishing between these interpretations would require detecting wavelength-dependent ring scattering in reflected light, potentially including polarimetric signatures as predicted by \citet{Veenstra2025}.

More generally, a planet with an extended atmosphere and an inclined ring system would appear doubly inflated, with both effects contributing to the observed transit depth. As revealed by our grid search, the inferred bulk densities can increase by modest factors of $\sim 2$--$5$ in some local probability maxima, which still leaves the planets substantially less dense than typical sub-Neptunes. However, we also find equally valid configurations with significantly smaller planetary radii where the bulk density increases by factors of $\sim 15$--$100$, placing them well within the typical density range of volatile-rich or even rocky planets. This multimodal density landscape suggests that rings, if present, could contribute to the super-puff phenomenon depending on the exact system geometry.

An important advantage of asterodensity profiling is its computational efficiency, as deriving $\rho_{\star,\mathrm{obs}}$ requires only the fundamental transit observables $(\delta, T_{14}, T_{23}, P)$. This makes the \PhotoRing\ diagnostic readily applicable to the full population of super-puff planets discovered by \textit{Kepler} and K2 without dedicated ring-transit modeling. While multi-planet architectures offer a distinct advantage by allowing the cross-testing of inter-planetary inconsistencies in \(\rho _{\star ,\mathrm{obs}}\), single-planet systems require a far more exhaustive analysis due to the lack of baseline references for contrast. Nevertheless, any system displaying such inter-planetary discrepancies, similar to those identified here, remains a prime target for detailed ring retrieval analyses. The case of \exoplanet{HIP-41378}{f} already provides the closest precedent, as its super-puff-like density was reconciled with an inclined ring system capable of inflating the apparent transit radius \citep{Akinsanmi2020}. If \PhotoRing\ offsets are found to be common among super-puffs, this would support the idea that unresolved ring systems contribute systematically to their apparent radius inflation, with important implications for sub-Neptune occurrence rates and their theoretical formation models.

More broadly, the inference framework developed here, combining \photofit-based asterodensity profiling, a geometric ringed-planet transit model, and Bayesian retrieval, is readily transferable to other systems requiring only posterior distributions for the transit observables and an independent estimate of $\rho_{\star,\mathrm{true}}$. Its main limitation is the photometric cadence required to resolve ingress and egress morphology. In the near term, existing JWST observations offer the most promising opportunity to test this scenario: JWST can directly constrain $b$ from resolved contact slopes, and its spectroscopic capabilities (along with ARIEL's), can simultaneously probe its atmospheric properties. The upcoming high-precision observations from PLATO$~$\citep{Matuszewski2023} for thousands of cold Jupiters will also enable systematic searches for ring signatures, improving TTV-based mass constraints, and identifying planets with anomalously low bulk densities.

\section*{Data Availability}

All the data necessary to reproduce the results and recreate the figures in this work—including the {\tt Jupyter} notebooks used for model training and evaluation—are publicly accessible at \url{https://github.com/seap-udea/PRisma} (PhotoRing inference and modeling algorithm). We anticipate that this project will provide a foundation for future implementations of the PR inference pipeline proposed in this paper. Contributions from the community are encouraged. 

\begin{acknowledgments}

This work has been possible due to the availability of open-source, general-purpose {\tt Python} packages, including 
\  {\tt Matplotlib} \citep{Matplotlib2007}, {\tt Numpy} \citep{Numpy2020}, {\tt Scipy} \citep{virtanenSciPy10Fundamental2020}, {\tt mpi4py} \citep{dalcin2021mpi4py} and {\tt pandas} \citep{Pandas2010}, \texttt{dynesty} \citep{Speagle2020} and \texttt{emcee} \citep{ForemanMackey2013}. JAAM acknowledges support from the Macquarie University Research Fellowship (MQRF).

\end{acknowledgments}

\newpage
\appendix

\section{Analytical Derivation of Ring Contact Positions using the Support Function}
\label{sec:appendix_support_function}

\subsection{Error Analysis of the 2015 Analytical Formula}

In the original formulation of the \PhotoRing\ effect by \citet{Zuluaga2015}, calculating the exact temporal contact points between a stellar limb and a projected exoplanetary ring requires solving a non-linear system of trigonometric equations. The previous analytical approximation assumed a highly elongated ellipse ($A \gg B$) and small contact angles, leading to geometrical artifacts at high projected tilts ($\theta_R \gtrsim 60^\circ$) and high inclinations.

\begin{figure*}[h!]
    \centering
    \includegraphics[width=0.48\textwidth]{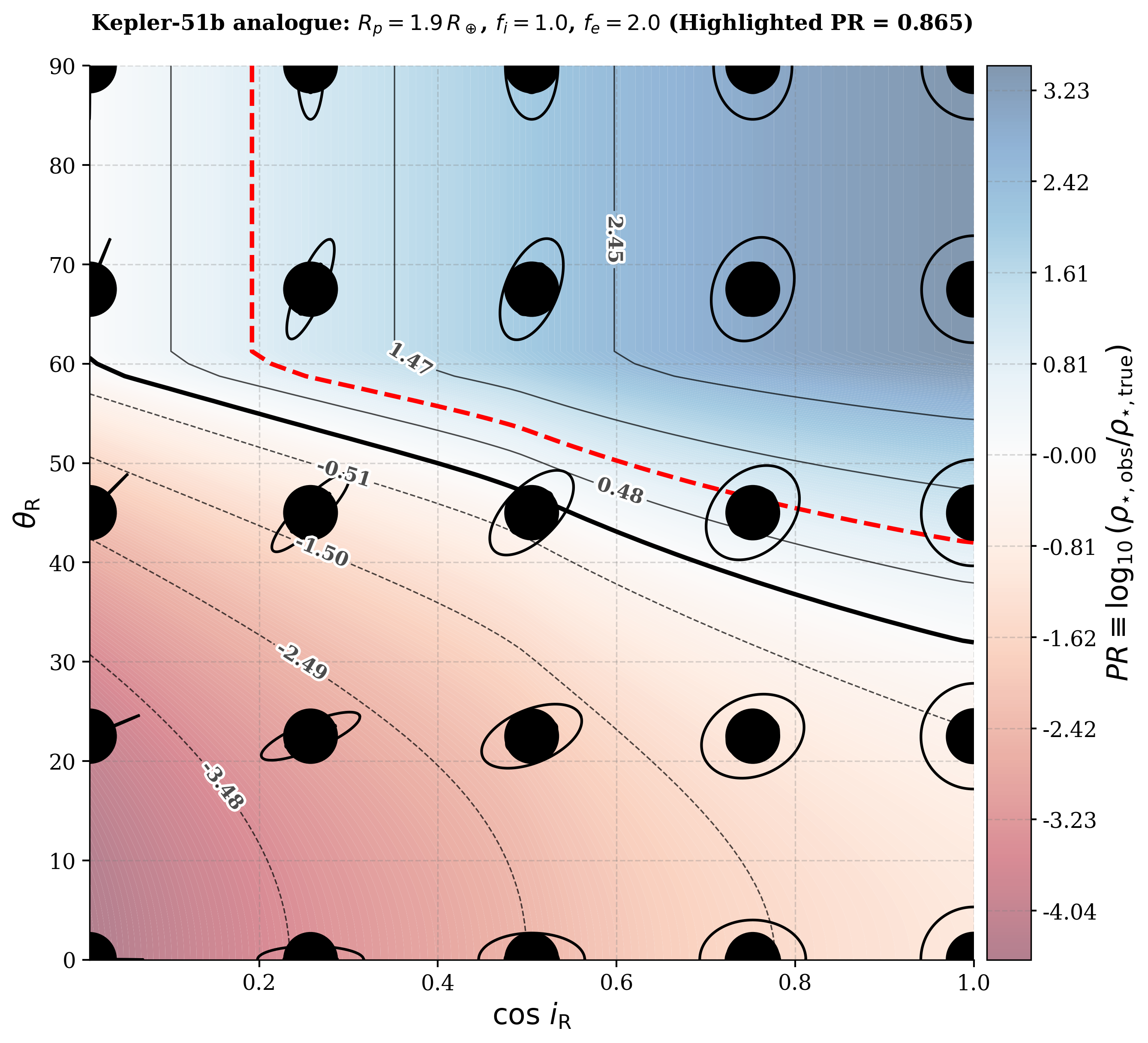}
    \includegraphics[width=0.48\textwidth]{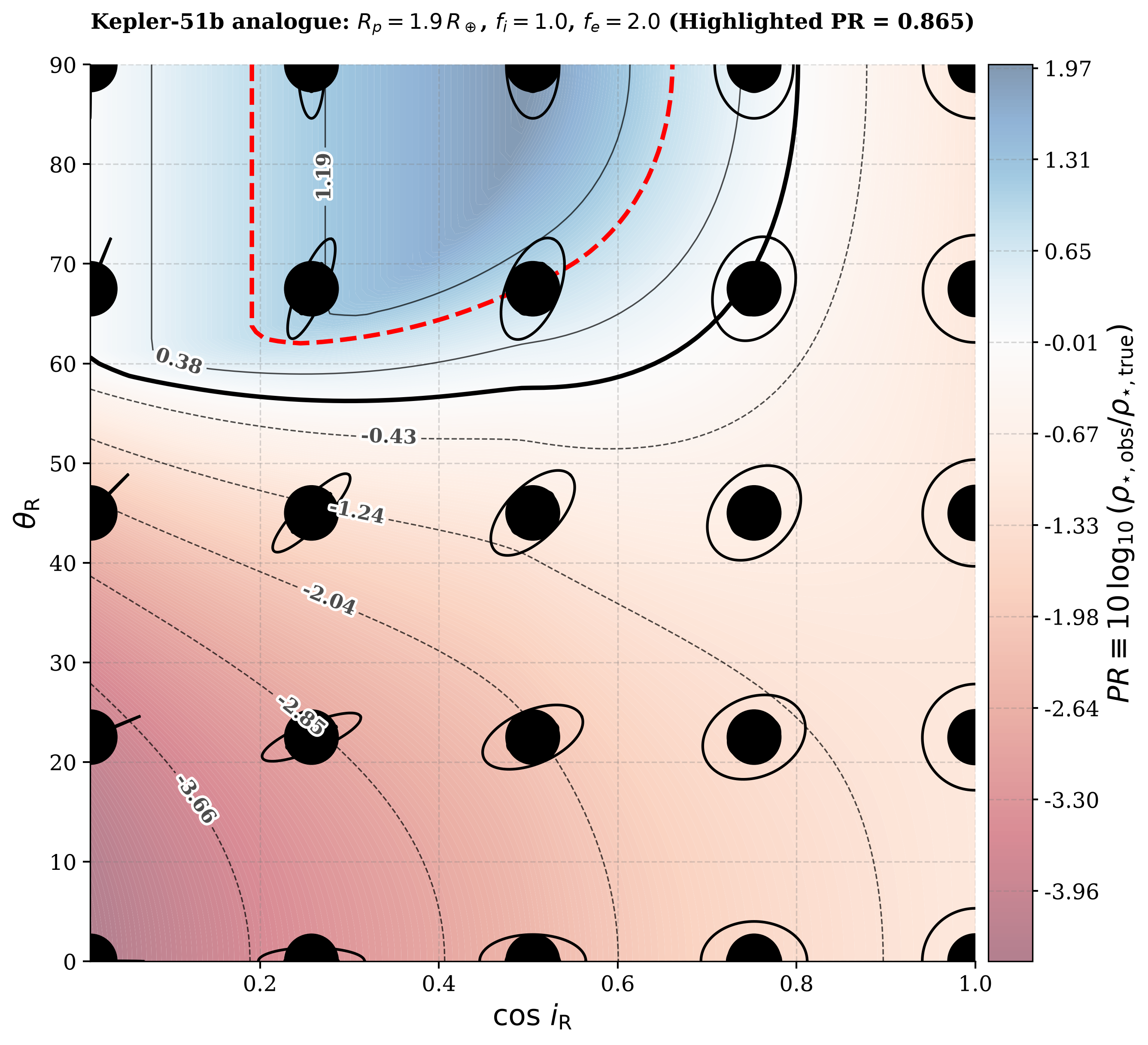}
    \caption{Comparison of the \PhotoRing\ effect contours for \exoplanet{Kepler-51}{b} using the original \texttt{exorings} analytical approximation from \citet{Zuluaga2015} (left) and the precise semi-analytical model \texttt{geotrans} (right). The original model exhibits geometrical artifacts at certain high projected tilts and inclinations. You can compare the panel on the right with \autoref{fig:photoring_contour}, which was obtained using the refined analytical model presented in this appendix.}
    \label{fig:pr_contour_comparison}
\end{figure*}

To illustrate this, \autoref{fig:pr_contour_comparison} compares the \PhotoRing\ effect contours generated by the precise semi-analytical model \texttt{geotrans} with those produced by the original \texttt{exorings} analytical approximation from \citet{Zuluaga2015}. As shown, the original formulation breaks down in certain inclination ranges, creating unphysical artifacts.

To provide a robust, exact-like analytical solution suitable for computationally intensive sampling (e.g., Nested Sampling), we reframe the contact geometry using the Minkowski Addition and the Support Function of a convex body \citep{schneider1993convex}.

\subsection{Theoretical Framework: The Support Function}

Geometrically, the contact between the stellar disk (a circle of radius $R_* = 1$) and the projected ring (an ellipse with semi-axes $A, B$ rotated by $\theta_R$) occurs when the distance between their centers equals the sum of their effective radii in the direction of contact. 

For a convex shape, the distance from its center to a tangent line perpendicular to a given normal vector $\vec{n}$ is defined by its Support Function, $h(\vec{n})$. For an ellipse aligned with the coordinate axes, the support function is:
\begin{equation}
    h(\vec{n}) = \sqrt{A^2 n_x^2 + B^2 n_y^2}
\end{equation}

\begin{figure*}[ht!]
\centering
\includegraphics[width=0.8\textwidth]{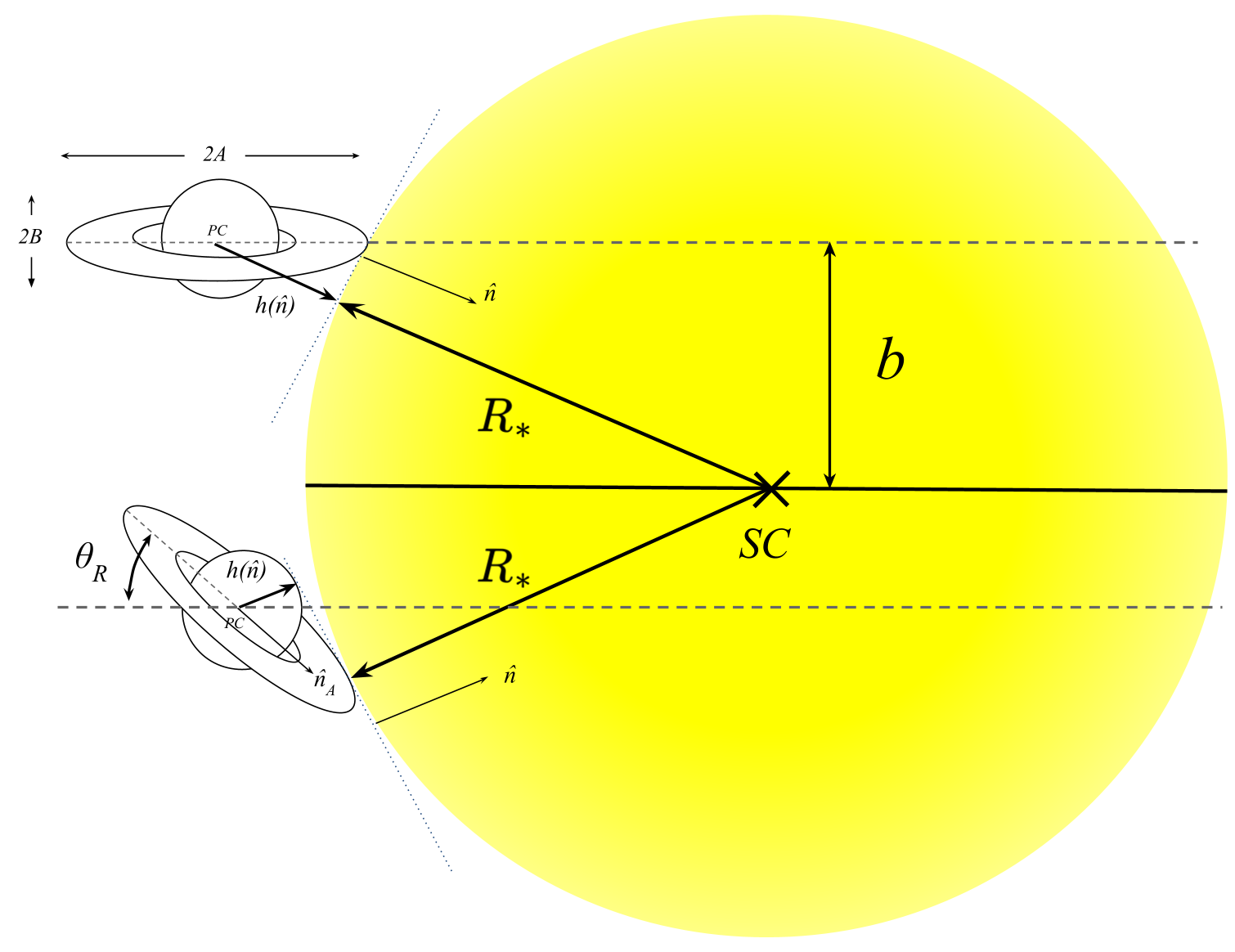}
\caption{Illustration of the quantities involved in the analytical model for two configurations: one where the rings are oriented coinciding with the coordinate system (left/top), and another where the rings are oriented with an inclination and projected tilt (right/bottom).}
\label{fig:analytical_model}
\end{figure*}

A graphical representation of these geometric quantities and the contact positions used in this derivation is provided in \autoref{fig:analytical_model}.

When the ellipse is rotated by an angle $\theta_R$ (the projected ring tilt), the normal vector must be rotated into the ellipse's local frame. The unit vectors defining the major and minor axes are:
\begin{equation}
    \hat{u}_A = (\cos\theta_R, \sin\theta_R), \quad \hat{u}_B = (-\sin\theta_R, \cos\theta_R) \nonumber
\end{equation}

The directional effective radius of the rotated ring is then:
\begin{equation}
    h(\vec{n}) = \sqrt{A^2 (\vec{n} \cdot \hat{u}_A)^2 + B^2 (\vec{n} \cdot \hat{u}_B)^2}
\end{equation}

\subsection{The Unperturbed Normal Approximation}

To avoid solving for the exact contact normal iteratively, we make a highly accurate physical approximation: the normal vector at the point of contact is nearly identical to the normal vector if the planet were a spherical point-mass.

For a point-mass transit with an impact parameter $b$, the contact points on the stellar limb have the coordinates $y = b$ and $x = \pm\sqrt{1-b^2}$. Therefore, the normal vectors at the egress (trailing, $R$) and ingress (leading, $L$) contacts are simply:
\begin{eqnarray}
    \vec{n}_R &=& \left( +\sqrt{1-b^2}, b \right) \\
    \vec{n}_L &=& \left( -\sqrt{1-b^2}, b \right)
\end{eqnarray}

Let $x_0 = \sqrt{1-b^2}$. We can now evaluate the Support Function of the ring in these two specific directions to obtain the effective contact radii, $h_R$ and $h_L$.

Projecting $\vec{n}_R$ onto the ellipse axes yields the egress effective radius:
\begin{eqnarray}
    \vec{n}_R \cdot \hat{u}_A &=& x_0 \cos\theta_R + b \sin\theta_R \nonumber \\
    \vec{n}_R \cdot \hat{u}_B &=& -x_0 \sin\theta_R + b \cos\theta_R \nonumber
\end{eqnarray}
Squaring and substituting into the support function gives:
\begin{eqnarray}
\label{eq:h_R}
    h_R^2 &=& A^2 (x_0 \cos\theta_R + b \sin\theta_R)^2 \nonumber \\
          & & +\, B^2 (b \cos\theta_R - x_0 \sin\theta_R)^2
\end{eqnarray}

Similarly, for the ingress effective radius ($h_L$):
\begin{eqnarray}
    \vec{n}_L \cdot \hat{u}_A &=& -x_0 \cos\theta_R + b \sin\theta_R \nonumber \\
    \vec{n}_L \cdot \hat{u}_B &=& x_0 \sin\theta_R + b \cos\theta_R \nonumber
\end{eqnarray}
Yielding:
\begin{eqnarray}
\label{eq:h_L}
    h_L^2 &=& A^2 (-x_0 \cos\theta_R + b \sin\theta_R)^2 \nonumber \\
          & & +\, B^2 (b \cos\theta_R + x_0 \sin\theta_R)^2
\end{eqnarray}

\subsection{Final Contact Equations}

With the effective geometrical radii $h_R$ and $h_L$ defined, the complex ring behaves instantaneously like a sphere of radius $h_{L,R}$. We substitute these directional radii into the standard spherical contact equations. The four contact positions along the $x$-axis are therefore given exactly by:

\noindent \textbf{Ingress Contacts (Left):}
\begin{eqnarray}
    x_{R,1} &\approx& -\sqrt{(1 + h_L)^2 - b^2} \quad \text{(First contact, external)} \\
    x_{R,2} &\approx& -\sqrt{\max(0, [1 - h_L]^2 - b^2)} \quad \text{(Second contact, internal)}
\end{eqnarray}

\noindent \textbf{Egress Contacts (Right):}
\begin{eqnarray}
    x_{R,3} &\approx& +\sqrt{\max(0, [1 - h_R]^2 - b^2)} \quad \text{(Third contact, internal)} \\
    x_{R,4} &\approx& +\sqrt{(1 + h_R)^2 - b^2} \quad \text{(Fourth contact, external)}
\end{eqnarray}

The argument of the square root for internal contacts ($x_{R,2}, x_{R,3}$) is bounded to $\ge 0$ to properly handle grazing configurations where a full interior transit does not occur.

\bibliography{references}{}
\bibliographystyle{aasjournalv7}

\end{document}